\documentclass[
    aps,
    prd,
    reprint,
    amsmath,amssymb,
    groupedaddress,
    twocolumn,showpacs,
    ]{revtex4-2}
\usepackage{graphicx}
\usepackage{dcolumn}
\usepackage{bm}
\usepackage[mathlines]{lineno}
\usepackage{multirow}
\usepackage{comment}
\usepackage{enumerate}
\usepackage{color}
\usepackage{amsmath}

\newcommand{\hatC}{\hat{C}}
\newcommand{\mD}{{\mathcal{D}}}
\newcommand{\mF}{\mathcal{F}}
\newcommand{\fe}{{ f_{\rm e}} }
\newcommand{\hatF}{\hat{F}}

\newcommand{\la}{{\langle}}
\newcommand{\ra}{{\rangle}}
\newcommand{\weylP}{{\Phi+\Psi}}

\begin{document}


\title{Cavendish experiment with fast radio bursts on cosmological scales}


\author{Shuren Zhou $^{1,2,3,4}$}
 \email{zhoushuren@sjtu.edu.cn}
\author{Pengjie Zhang $^{2,1,3,4}$}
 \email{zhangpj@sjtu.edu.cn}
\affiliation{
$^{1}$ Tsung-Dao Lee Institute, Shanghai Jiao Tong University, Shanghai 200240, China  \\
$^{2}$ School of Physics and Astronomy, Shanghai Jiao Tong University, Shanghai 200240, China  \\
$^{3}$  Key Laboratory for Particle Astrophysics and Cosmology (MOE)/Shanghai Key Laboratory for Particle Physics and Cosmology, Shanghai 200240, China   \\
$^{4}$  State Key Laboratory of Dark Matter Physics, Shanghai 200240, China  
}


\date{\today}


\begin{abstract}
A key measure of gravity is the relation between the Weyl potential $\Psi+\Phi$ and the matter overdensity $\delta_m$, encapsulated as an effective gravitational constant $G_{\rm light}$ for light motion. Its value, along with possible spatial and temporal variations, is essential for probing physics beyond Einstein gravity.  However, the absence of an unbiased proxy for $\delta_m$ prevents the direct measurement of $G_{\rm light}$. In this work, we show that within a theoretical framework respecting the weak equivalence principle, the dispersion measure (DM) of localized fast radio bursts (FRBs) serve as a good proxy for $\delta_m$. We further propose an FRB-based estimator $F_G$ to directly measure $G_{\rm light}$, combining galaxy-DM of localized FRBs and galaxy-weak lensing cross-correlations. With a conservative cut $k\leq 0.1\, h/{\rm Mpc}$, the measurement can achieve a precision of $\lesssim 10\% \sqrt{10^5/N_{\rm FRB}}$ over 10 equal-width redshift bins at $z\lesssim 1$. The major systematic error, arising from the clustering bias of electrons traced by the FRB DM, remains subdominant at the $5\%$ level. It can be further mitigated to the $\lesssim 1\%$ level, based on the gastrophysics-agnostic behavior that the clustering bias of total baryons (ionized diffuse gas, stars, neutral hydrogen, etc) approaches unity at sufficiently large scales. Therefore, FRBs shed light on gravitational physics across spatial and temporal scales spanning 20 orders of magnitude. 
\end{abstract}


\maketitle


\section{Introduction}
General Relativity (GR) is a cornerstone of the standard cosmology. This makes cosmological tests of GR  crucial, in particular for understanding the observed cosmic acceleration  \cite{weinberg2013observational, kamionkowski2023hubble} and distinguishing between dark energy (DE) and modified gravity (MG) \cite{jain2008observational,clifton2012modified, ezquiaga2018dark, joyce2015beyond, koyama2016cosmological, ishak2019testing, ferreira2019cosmological}. Recently reported evidence for dynamical dark energy from the baryonic acoustic oscillation (BAO) analysis by the Dark Energy Spectroscopic Instrument (DESI) \cite{karim2025desi, lodha2025extended, gu2025dynamical} made this task even more important and urgent. 

The impact of MG on the large-scale structure (LSS) of the universe can be parameterized by two parameters $G_{\rm light}$ and $\eta\equiv \Phi/\Psi$, or other equivalent parameterizations (e.g., \cite{zhang2007probing, zhao2009searching, levon2016can}). 
Here $G_{\rm light}$ is the effective gravitational constant in the relation between the Weyl potential $\Psi+\Phi$ that light senses and the matter overdensity $\delta_m$, with convention $d\tau^2=(1+2\Psi)dt^2-a^2(1-2\Phi)d{\bf x}^2$. MG models in general lead to $G_{\rm light}\neq G$, $\eta\neq 1$, or both. Recent observations have started to put useful constraints on these parameters through full-shape data analysis and joint fitting together with other parameters \cite{ade2016planck, alam2021completed, ishak2024modified}.
Meanwhile, efforts aiming at measuring these parameters with less dependence on LSS modeling are actively underway. One example is the $E_G$ estimator \cite{zhang2007probing}, which combines galaxy-weak lensing and galaxy-velocity cross-correlations into a single measurement of $E_G\propto (G_{\rm light}/G)/f$, where $f\equiv d\ln \delta_m/d\ln a$. $E_G$ and its extensions have been implemented across multiple surveys \cite{reyes2010confirmation, pullen2016constraining, amon2018kids, singh2019probing,ishak2019testing, zhang2021testing, wenzl2024constraining, wenzl2025atacama, jullo2019testing, blake2020testing, alam2017testing, blake2016rcslens, rauhut2025testing, li2025testing}. Nonetheless, their measurement relies on modeling of  redshift space distortion and the linearized continuity equation, whose impact will eventually become significant. 

The bottleneck in $G_{\rm light}$ measurement is to probe $\delta_m$, since we can no longer infer $\delta_m$ from weak lensing as in the GR framework. We advocate that  this issue will eventually be resolved by localized fast radio bursts (FRBs, \cite{ioka2003cosmic, zhang2023physics}). The weak equivalence principle implies that dark matter and baryonic matter share the same spatial distribution on $\gtrsim 10$ Mpc scales where gravity dominates over all other forces, namely $\delta_m=\delta_b$. 
Meanwhile, FRBs probe the distribution of free electrons in ionized diffuse gas through the dispersion measure (DM), 
\begin{equation} \label{equ:D_express}
\mD = {3H_0^2\over 8\pi G} {\Omega_{b0} \over m_p} \int d\chi\, a^{-1}\fe \left(1+\delta_e\right) \ .
\end{equation}
Here $\delta_e$ is the electron density fluctuation along the radial distance $\chi$, and $\fe \equiv f_{\rm HII} + {1\over 2}f_{\rm HeIII}$ is the ionization fraction. Since these electrons represent the majority  ($\gtrsim 90\%$) of cosmic baryons \cite{macquart2020census,connor2025gas}, we expect $\delta_e\simeq \delta_b$. 
The accuracy of this approximation can be further improved to the 1\% level through a mitigation method proposed in this work. Therefore, DM of FRBs serves as an unbiased tracer of $\delta_m$. 

To infer $\delta_m$ robustly, the localization of FRBs is demanded to identify the host galaxy redshifts. This redshift information enables the isolation of the intergalactic medium (IGM) contribution $\mD$ from the host galaxy DM, using the galaxy-DM cross-correlation statistics. 
Meanwhile, despite that the physical origin of FRBs is not settled \cite{pen2018nature}, the event rate of FRBs is significant \cite{fialkov2017fast}. 
The planned radio arrays such as DSA-2000 \cite{hallinan2019astro2020} are expected to detect $\sim 10^{4}$ localized FRBs each year, and BURSTT \cite{lin2022burstt} is also optimized to detect and localize a large sample of FRBs. A sample of $\sim 10^5$ localized FRBs up to $z\sim 1$ for LSS statistics is achievable in the foreseeable future. 


In this work, we design a cosmological Cavendish experiment through $F_G$, a FRB-based estimator of gravity. We demonstrate that it is capable of measuring $G_{\rm light}$ with $\lesssim 10\% \,\sqrt{10^5/N_{\rm FRB}}$ precision and $\sim 1\%$ accuracy over many redshift bins, far exceeding the existing constraints. This measurement will put a stringent constraint on MG models. For instance, a detection of $G_{\rm light}\neq G$ would rule out GR and $f(R)$ gravity, as both predict $|G_{\rm light}/G-1|\ll 1$ \cite{zhang2006testing, ishak2019testing}, and the whole Hordenski scalar-tensor theory would be ruled out when combined with $c_{\rm GW}=c$ verified by GW170817 \cite{levon2016can, abbott2017gw170817, ezquiaga2017dark, creminelli2017dark, baker2017strong}. 

\section{The FRB-based gravity estimator }
The $F_G$ estimator combines three tracers $X\in\{\Delta_g, \kappa, \mD\}$: the galaxy surface overdensity $\Delta_g$, the lensing convergence $\kappa$, and the DM $\mD$ of localized FRBs. 
They are related to the underlying 3D overdensity $Y\in\{\delta_g,\, \nabla^2(\weylP),\, \delta_e\}$ by $X(\hat{n})=\int Y(\hat{n},\chi)\, W_X(\chi)d\chi$, where $W_X$ is the kernel function. 
For $\Delta_g$ of a given galaxy redshift bin, the cross-correlation $\langle \Delta_g\kappa\rangle$ and $\langle \Delta_g\mD \rangle$  isolate $\weylP$ and $\delta_e$ within the bin respectively. The Weyl potential is related to $\delta_m $ through 
\begin{equation}
\nabla^2\left(\Phi+\Psi\right) = 8\pi G_{\rm light}a^2\bar\rho_m {\Delta}_m   \;. 
\end{equation}
Here ${\Delta}_m$ is the gauge-invariant matter density contract, which reduces to ${\Delta}_m=\delta_m$ in comoving-synchronous gauge \cite{jeong2012large}. 
Therefore
\begin{equation}
    \la\Delta_g\kappa\ra \propto G_{\rm light} \la\Delta_g\mD\ra\ .
\end{equation} 
Then, the $F_G$ estimator is defined in Fourier space by 
\begin{equation}\label{equ:FG-ell}
\hatF_G \equiv  \hat\mF\,  { \hatC^{g\kappa}_\ell \over \hatC^{g\mD}_\ell } 
\;,
\end{equation}
which is the ratio of $\Delta_g$-$\kappa$ and $\Delta_g$-$\mD$ angular power spectra, and adopts the definition $\la A_\ell B_{\ell'}\ra = \delta^D_{\ell\ell'} C^{AB}_\ell$. The fields $\kappa$ and $\mD$ are integrated over the entire line of sight, while the galaxy clustering with redshift information enables the tomographic slicing of the radial projection. The normalization $\mF$ is chosen such that the expectation value is 
\begin{equation}
    F_G=\frac{G_{\rm light}}{G}\ .
\end{equation}
Note that the ratio $G_{\rm light}/G$ is often parameterised as $\Sigma$ or $1+\Sigma$ in literature. The applications of Eq.~(\ref{equ:FG-ell}) are feasible for both narrow and wide redshift bins, as long as $\mF$ is defined correspondingly. To maintain the redshift resolution, we choose galaxy samples with a narrow width $\Delta z \ll 1$ and denote the centered mean redshift as $z_g$. Meanwhile, Eq.~(\ref{equ:FG-ell}) is applicable regardless of whether the Limiber approximation holds, while we adopt it for brevity. $\mF$ is then given by 
\begin{align}
\mF 
&\equiv  { \langle \hatC^{g\mD}_\ell \rangle\over \langle \hatC^{g\kappa}_\ell \rangle} \bigg|_{\rm GR}   
\simeq { W_\mD(z_g) \over W_\kappa(z_g) } {P_{ge}\over P_{gm} }     \nonumber \\
&= {1\over 4\pi G m_p} {\Omega_{b0}\over\Omega_{m0}} { N_{\mD}(z_g) \over \chi_g\,N_{\kappa}(z_g) } \fe\, b_e    \;,   \label{equ:mF}
\end{align}
where $b_e\equiv P_{me}/P_{mm}$ is the electron bias. 
The kernel functions are specifically $W_\kappa(\chi) = {3\over 2}\Omega_{m0}H_0^2\, a^{-1} \chi N_\kappa(z) $  and $W_\mD(\chi) = {3H_0^2\over 8\pi G} {\Omega_{b0} \over m_p} \, \fe\, a^{-1}\, N_\mD(z)$, 
where the integrations of the source distribution give 
$ N_{\mD}(z) = \int_{z}^\infty n_\mD(z') dz'$ and 
$ N_{\kappa}(z) = \int_{z}^\infty dz' n_\kappa(z') \left( 1-{\chi/\chi'}\right)  \,$, 
respectively \cite{zhang2023physics, kilbinger2015cosmology}. 
Note that the ensemble average of the ratio does not equal the ratio of the ensemble averages, in particular if the denominator has large statistical errors. So in practice, $G_{\rm light}$ is obtained by fitting two cross power spectra against the proportionality relation $C^{g\kappa}_\ell \propto G_{\rm light} C^{g\mD}_\ell$ through the ratio measurement method \cite{sun2023unbiased}, which yields unbiased results even if the denominator $\hat{C}^{g\mD}_\ell$ is noisy.

The $F_G$ estimator not only directly measures $G_{\rm light}$, but also enjoys many complementary features compared to previous tests of gravity. 
For the tomography of redshift evolutions, $F_G$ relies solely on the known redshift distribution of galaxies, which enables its application to imaging surveys with larger galaxy samples.
Upon the clustering bias, the galaxy deterministic bias from cross-correlation generally differs from the value inferred from auto-correlation, due to the non-Poisson nature of stochasticity in galaxy clustering \cite{bonoli2009halo, seljak2009suppress, hamaus2010minimizing, cai2011optimal, liu2021biased, zhou2024parametrization}. It leads to a suppression of the ratio between cross- and auto-correlation, and potentially underestimates the estimator like $E_G$ 
\footnote{
For illustration, we consider the estimator $\hat{E}_G\propto \hatC^{g\kappa}/(\beta\hatC^{gg})$ in Ref.~\cite{pullen2015probing} designed for projected fields. 
In the approximation of narrow redshift bin, there are $\hatC^{g\kappa}\propto P_{gm}=b^D P_{mm}$ and $\hatC^{gg}\propto P_{gg} = (b^S)^2 P_{mm}$. 
The deterministic bias $b^D$ differs from the stochastic bias $b^S$ in the presence of galaxy stochasticity. 
Meanwhile, the redshift-space distortion parameter is $\beta = f/b^D$, leading to $\hat{E}_G\propto (b^D/b^S)^2 = r^2_{gm} $.
Consequently, the $E_G$ is suppressed by a factor $r^2_{gm}<1$, which is the cross-correlation coefficient between the underlying matter and the galaxy clustering. 
}. 
In contrast, $F_G$ relies exclusively on cross-correlations. Furthermore, $F_G$ avoids the contamination of the DM by the host galaxy or the Milky Way, which is typically removed with large uncertainty and model dependence in the full-shape analysis of DM.

\section{Detection Significance}
The uncertainty of the estimator Eq.~(\ref{equ:FG-ell}) consists of two contributions, the angular power spectrum measurement $\sigma_{C_\ell}^2$ and the overall amplitude estimation $\sigma_\mF^2$, i.e., 
$ \sigma_{\ell}^2 = \sigma_{C_\ell}^2 + \left(F_G/\mF\right)^2\, \sigma^2_\mF$.
We have assumed the statistical errors of $\hatC_\ell^{gX}$ and $\hat\mF$ are uncorrelated, since in principle, they are independent measurements from different probes. 
The former part is estimated using the Gaussian field approximation, 
\begin{equation}\label{equ:variance-FG-Cell}
{\sigma_{C_\ell}^2\over F_G^2} = {1\over (2l+1)f_{\rm sky}} \left[
{ \hatC^{gg}_\ell \hatC^{\kappa\kappa}_\ell \over  \left(\hatC^{g\kappa}_\ell \right)^2 } + 
{ \hatC^{gg}_\ell \hatC^{\mD\mD}_\ell  \over  \left(\hatC^{g\mD}_\ell \right)^2 } - 2
{ \hatC^{gg}_\ell \hatC^{\kappa\mD}_\ell  \over  \hatC^{g\kappa}_\ell C^{g\mD}_\ell  }
\right]     \;.
\end{equation}
Here, the shot noise is included in the auto-power spectrum for variance estimation 
\footnote{
The shot noise contribution $\sigma^2_{\rm DM} = \sigma^2_{\rm host} + \sigma^2_\mD + \sigma^2_{\rm MW}$ in the power spectrum of DM is not properly accounted in some forecast works, such as Ref.~\cite{neumann2024fast, shirasaki2022probing}, where they only consider the subdominant host-galaxy contribution $\sigma^2_{\rm host}$ and result in over-optimistic results. 
The cosmic DM $\sigma_\mD^2$ in high redshift FRB can be significantly larger than the host-galaxy contribution. 
Therefore, the shot noise can dominate the measurement $\hatC^{\mD\mD}_\ell = C^{\mD\mD}_\ell + \sigma^2_{DM}\,/\bar{n}_{\rm FRB}$ on scale $\ell\gtrsim 10\sim 100$ for typical FRB number $10^4\sim 10^5$ in usual estimation \cite{madhavacheril2019cosmology, sharma2025probing}. 
}.
To reduce the fluctuation in power spectra and to perform a scale-independent null test of GR, we can combine all available $\ell$-modes to construct the $F_G$ estimator, 
\begin{equation}\label{equ:FG-wei}
\hatF_G =  \hat\mF\, 
{ \sum_\ell\, w_\ell\, \hatC^{g\kappa}_\ell  \over  \sum_\ell\, w_\ell\, \hatC^{g\mD}_\ell } 
\;,
\end{equation}
where the minimum variance weight is $w_\ell = \left(C^{g\mD}_\ell \, \sigma^2_{C_\ell}\right)^{-1}$, and the corresponding estimator variance is
$\sigma^2_{F_G} = \left( \sum_\ell 1/\sigma^2_{C_\ell} \right)^{-1} + \left(F_G/\mF\right)^2\,\sigma^2_\mF$.

\begin{figure}
\includegraphics[width=\columnwidth]{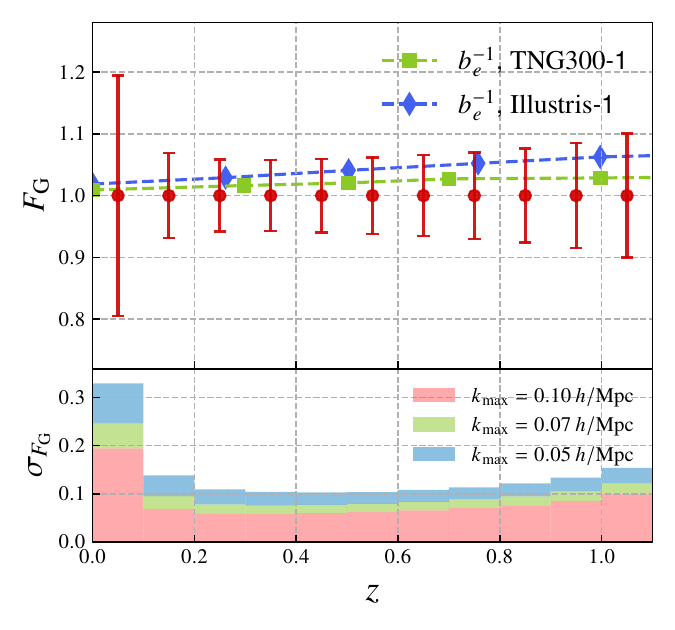}
\caption{ \label{fig:FG} 
Statistical errors of the tomographic $F_G$ measurement, assuming the fiducial values $F_G=1$, i.e., $G_{\rm light}=G$. 
The \textit{top panel} shows the forecast combining all scales up to $k_{\rm max}= 0.1\, h/{\rm Mpc}$ to estimate uncertainties, while the \textit{bottom panel} shows the uncertainties with varying $k_{\rm max}$. 
The $F_G$ measurement requires three data sets: a galaxy catalog chosen as a DESI-like catalog, a weak lensing catalog chosen as a Rubin-like shear catalog, and  a DM catalog of localized FRBs. The limiting factor is the number of localized FRBs, chosen as $N_{\rm FRB}=10^5$. 
Additionally, we present the reciprocal of electron bias $b_e^{-1}$ measured in simulations TNG300-1 ({green dashed}) and Illustris-1 ({blue dashed}). If uncorrected, it induces a systematic shift in $F_G$ at $1\%\sim 5\%$ level, which remains subdominant to statistical error. 
}
\end{figure}

We present the forecast of the detection significance of cross-correlating DESI bright galaxy (BGS, $0<z<0.4$) and luminous red galaxy (LRG, $0.4<z<1.1$), with cosmic shear detected by the Vera C. Rubin Observatory survey \cite{ivezic2019lsst}, and with DM from well-localized FRB samples. The galaxy number density and galaxy clustering bias are estimated using the complete DESI BGS and LRG samples \cite{adame2024validation}.
Both source distributions of backlight $X\in\{\kappa,\mD\}$ are modeled as $n_X(z)\propto z^2\,e^{-\alpha z}$ with $\alpha=2.5$, and cover a sky fraction of $f_{\rm sky}=0.34$ \cite{chang2013effective, ivezic2019lsst, rafiei2021chime, neumann2024fast}.  
For shear samples, we assume the surface number density of $36\; {\rm arcmin}^{-2}$, the shape noise of $\sigma_\epsilon = 0.3$, and the photometric redshift scatter of $\sigma_{pz}=0.02$ \cite{chang2013effective, ivezic2019lsst}. The estimation of $F_G$ detection significance is insensitive to these survey specifications adopted, since the limiting factor is the total number of localized FRBs. 

For a redshift bin $[z_1, z_2]$, we cross-correlate $\Delta_g$ with $\kappa$ sources at $z > z_2+\sigma_{pz}$ to avoid the contamination from intrinsic alignment. We also cross-correlate $\Delta_g$ only with $\mD$ sources at $z > z_2$ to prohibit the systematic impact of host galaxy DM. Nonetheless, the host galaxy DM contributes random noise in the cross-correlation measurement, and it is taken into account as a shot noise of $\sigma_{\rm host} = 100 {\rm\,pc\,cm^{-3}}$. 
As a simplified case shown in Fig.~\ref{fig:FG}, we assume perfect knowledge about the normalization $\mF$ and account only for the uncertainty $\sigma_{C_\ell}$, i.e., setting $\sigma_\mF=0$. 

The fiducial estimation is restricted to linear scales $k\leq 0.1\,h/{\rm Mpc}$, where baryonic feedback on the clustering is roughly negligible \cite{schaller2025flamingo, siegel2025suppression}. 
At low redshift $z\lesssim 0.1$, the measurement is therefore subject to cosmic volume. At higher redshift, it is primarily limited by the number of FRBs; for instance, shot noise overwhelms $C^{\mD\mD}_\ell$ signal on $\ell\gtrsim 50$ at $z\simeq 1$ with $N_{\rm FRB}=10^5$. 
In the region where shot noise dominates, the statistical uncertainty in $F_G$ is 
\begin{equation}
\sigma_{F_G}   \simeq  \, 0.1\,
\left( \sigma_{\rm DM}\over 200\, {\rm pc\,cm^{-3}} \right)
\left( N_{\rm FRB}\over 10^5 \right)^{-\frac{1}{2}}   \;,
\end{equation}
where $\sigma_{\rm DM}$ is the standard deviation of the observed DM  
\footnote{
We neglect the sub-dominating Milky Way contribution, and integrate the cosmic contribution $\sigma_\mD$ up to $\ell=3000$. 
}. Even upon a moderate estimation of $N_{\rm FRB}=10^5$, we achieve $\sigma_{F_G}\sim 8\%$ across $10$ redshift bins (top panel of Fig.~\ref{fig:FG}), corresponding to an overall precision of $2\%$ over $0\lesssim z\lesssim 1$.
With more conservative scale cuts, such as $k\leq 0.07\,h/{\rm Mpc}$ or $0.05 \,h/{\rm Mpc}$, the precision degrades slightly to the $10\%$ level per redshift bin at $z\gtrsim 0.1$. The overall precision remains at the precent level, reaching $2.8\%$ for $k_{\rm max}=0.07\,h/{\rm Mpc}$ and $3.7\%$ for $k_{\rm max}=0.05\,h/{\rm Mpc}$.


\begin{figure}
\includegraphics[width=\columnwidth]{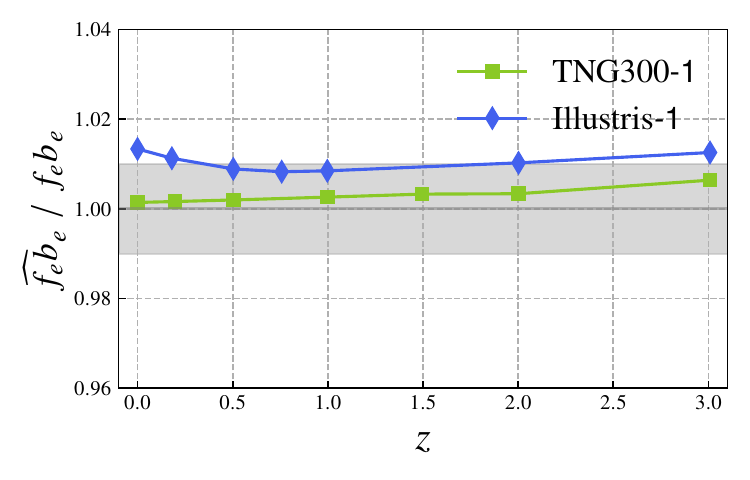}
\caption{ \label{fig:electron_bias} 
The residual systematic errors of the $F_G$ measurement after systematics mitigation, where $\widehat{f_eb}_e$ is estimated by Eq.~(\ref{equ:be_from_bb}) and $f_e b_e$ is the true value. 
The major systematic bias in $F_G$ arises from the determination of $f_eb_e$, where $b_e\neq 1$ as shown in Fig.~\ref{fig:FG}. The proposed Eq.~(\ref{equ:be_from_bb}) addresses this issue by expressing $f_eb_e$ in terms of stellar and neutral gas contributions, based upon the weak equivalence principle. 
Despite dramatically different strengths of AGN feedback adopted in simulations, both TNG300-1 (green line) and Illustris-1 (blue line) validate Eq.~(\ref{equ:be_from_bb}) to $1\%$ accuracy, demonstrating its insensitivity to these gastrophysics.
Therefore, $f_eb_e$ can be inferred using observations of stars and neutral gas, reducing the systematic errors in $F_G$ to the $\sim 1\%$ level. 
}
\end{figure}

\section{Mitigating potential systematics}
The uncertainty of the overall amplitude $\mF$ arises from two sources, cosmological parameters and gastrophysical effects. Utilizing the tight constraint of cosmological parameters from CMB observations \cite{aghanim2020planck, louis2025atacama} and BAO surveys \cite{alam2021completed, karim2025desi}, we can determine the cosmic geometric term ${\Omega_{b0}\over\Omega_{m0}} { N_{\mD}(z_g) \over \chi_g\,N_{\kappa}(z_g) }$ precisely. 
However, the electron bias $b_e$ arising from astrophysical processes is not a direct observable. 
We propose a solution exploiting the fact that in the late universe, the total baryonic matter is an unbiased tracer of the matter distribution on sufficiently large scales. Accordingly, the clustering bias of the total baryon is unity at redshifts $z\lesssim 2$ and scales $k\lesssim 0.1\,h/{\rm Mpc}$, provided that the non-gravitational effects are negligible above Mpc scales. By isolating all neutral gas baryons and stellar baryons from the total baryon budget, we can estimate the electron bias $b_e$ together with the ionized electron fraction $f_e$, 
\begin{equation}\label{equ:be_from_bb}
f_e b_e \simeq 
{ X_{\rm H} + {1\over 2} X_{\rm He}  \over  X_{\rm H} + X_{\rm He}  }
\left(  1 - f_{*} b_{*} 
- {X_{\rm H}+X_{\rm He}\over X_{\rm H} } f_{\rm HI} b_{\rm HI}  \right)   \;.     \\
\end{equation}
Here $b_i\equiv P_{im}/P_{mm}$ denotes the bias of $i$-species component, and $f_i\equiv\Omega_i/\Omega_b$ is the baryon mass fraction. $X_{\rm H}\simeq 0.76$ and $X_{\rm He}\simeq 0.24$ are mass abundances of hydrogen and helium elements. We only consider the significant baryon components, where $f_{*} b_{*}$ is the contribution of stars and stellar remnants, and $f_{\rm HI} b_{\rm HI}$ is the contribution of neutral hydrogen. The derivation of Eq.~(\ref{equ:be_from_bb}), along with the measurements of the mass fraction and clustering bias of these baryonic components in hydrodynamical simulations, is presented in the appendix. 

A major challenge in inferring $f_{*} b_{*}$ from observations is to convert galaxy luminosity into stellar mass, which is subject to the uncertainty in the stellar initial mass function. Another issue involves the fraction of stellar mass in faint galaxies, which limits the completeness of samples. 
However, the stellar censuses such as Gaia \cite{vallenari2023gaia, lutsenko2025counting} and ALMA \cite{pouteau2022alma, pouteau2023alma} are substantially enhancing our knowledge of nearby populations, and the constraints on the stellar distribution across cosmic time are also rapidly advancing. For instance, using imaging from DECaLS surveys, the stellar mass function is already extended into the $10^6\,M_\odot$ frontier \cite{xu2025pac, wang2025luminosity}, while Rubin will reach depths $2\sim 3$ magnitudes fainter \cite{ivezic2019lsst}. So resolving the stellar content is promising in the coming years. 
Upon the neutral gas contribution, the radio surveys such as CHIME \cite{amiri2022overview} and SKA \cite{braun2015advancing} are mapping the neutral hydrogen intensity across a wide redshift range through the surface brightness temperature of 21cm lines; for instance, the cross-correlations with LSS tracers have been conclusively detected \cite{masui2013measurement, cunnington2023h, amiri2023detection, amiri2024detection}. It enables the investigation of cold gas distribution and allow the precise measurement of $f_{\rm HI}b_{\rm HI}$ in the near future. 
Therefore, the determination of $f_eb_e$ is available through Eq.~(\ref{equ:be_from_bb}) combining the constraints from external probes \footnote{Besides those based upon Eq. ~(\ref{equ:be_from_bb}), the kinetic Sunyaev–Zel'dovich effect offers another potential pathway to constrain $b_e$ given its sensitivity to all free electrons in diffuse gas, yet its reconstruction such as four-point estimations depends on the template of LSS tracer velocities \cite{smith2017detecting, kumar2025electrons}. }, 
and thereby fully determines the value of $\mF$ factor.

Nevertheless, corrections from latter two terms in $f_eb_e\propto 1 - f_{*} b_{*} 
- (X_{\rm H}+X_{\rm He})/ X_{\rm H}\, f_{\rm HI} b_{\rm HI} \;$ is expected to be minor, as the universe is nearly fully ionized at low redshift $z \lesssim 2$ and their combined contribution amounts to merely $f_{*} + f_{\rm HI}\lesssim 10\%$. 
In the lowest order approximation, there are $f_e\simeq X_{\rm H} + {1\over 2} X_{\rm He}$ and $b_e\simeq 1$. 
Since ionized electrons in diffuse gas reside predominantly in the underdense regions of the cosmic web, leading to $b_e<1$, this rough approximation would result in an overestimation as $\hatF_G\propto b_e^{-1}$. 
To quantify the potential impact, we also present the $b_e$ measurement in Fig.~\ref{fig:FG}, using hydrodynamical simulations TNG300-1 from the IllustrisTNG project \cite{Springel_2017, Nelson_2017, Pillepich_2017, Naiman_2018, Marinacci_2018} and its predecessor Illustris-1 \cite{vogelsberger2014properties, vogelsberger2014introducing, genel2014introducing, sijacki2015illustris}. The strong baryon feedback in Illustris-1 yields better agreement with recent detections of Sunyaev-Zel'dovich effects by DESI tracers \cite{chen2023thermal, hadzhiyska2024evidence, guachalla2025backlighting}. 
Without any systematic mitigation, hydrodynamic simulations suggest that neglecting electron bias leads to systematic shifts of $\lesssim 3\%$ by TNG300-1 and $\lesssim 8\%$ by Illustris-1. 
This systematic effect remains subdominant compared to the DM shot noise with $N_{\rm FRB} = 10^5$. By employing Eq.~(\ref{equ:be_from_bb}) to infer the full $f_e b_e$, this subdominant systematic can be further reduced. As shown in Fig.~\ref{fig:electron_bias}, both TNG300-1 and Illustris-1 validate the accuracy of Eq.~(\ref{equ:be_from_bb}) to within $\lesssim 1\%$ across a wide redshift range $0 < z < 3$, despite these simulations employing different subgrid physics. Even upon an optimistic scenario of $N_{\rm FRB} = 10^6$, realizing $\sigma_{F_G}\sim 3\%$ for a redshift bin, the residual systematic at the sub-percent level is negligible.

Since the contributions from neutral gas and the stellar components are subdominant, their uncertainties are largely suppressed in the overall systematic mitigation of Eq.~(\ref{equ:be_from_bb}). The uncertainty in the estimation of $f_eb_e$ propagates to $F_G$ as $(\delta F_G /F_G)_{\rm sys} = \delta(f_eb_e)/f_eb_e$, where 
\begin{align}
{\delta(f_eb_e)\over f_eb_e} \simeq 
1.1\, {\delta(f_*b_*)\over f_*b_*} f_*b_* +  
1.5\, {\delta(f_{\rm HI}b_{\rm HI})\over f_{\rm HI}b_{\rm HI}} f_{\rm HI}b_{\rm HI}   \;.
\end{align}
In TNG300-1, the values of $f_*b_*$ and $f_{\rm HI}b_{\rm HI}$ are approximately $0.01 \sim 0.05$. To achieve $\lesssim 1\%$ uncertainty in $f_eb_e$ and hence in $F_G$, the relative uncertainties of ${\delta(f_*b_*)/f_*b_*}$ and ${\delta(f_{\rm HI}b_{\rm HI}) /f_{\rm HI}b_{\rm HI}}$ are required to be constrained at the $10\% \sim 50\%$ level individually.
In Illustris-1, $f_*b_*$ and $f_{\rm HI}b_{\rm HI}$ are larger, at approximately $0.05 \sim 0.1$. In this case, their fractional uncertainties must be controlled below the $5\% \sim 10\%$ level to maintain sub-percent accuracy in the systematic mitigation.

\section{Discussions and Conclusions}

In this work, we demonstrate that the DM of localized FRBs is a good proxy for $\delta_m$, combining the facts that DM is a direct probe of baryon distribution $\delta_b$ and the equivalence principle implies $\delta_b=\delta_m$ on large scales. We further propose the FRB-based estimator $F_G$ for the cosmological test of gravity theory that retains the weak equivalence principle, where $F_G$ directly measures $G_{\rm light}\propto (\weylP)/\delta_m$ across tomographic redshifts. 
The major systematic impact from the electron bias is subdominant relative to the DM shot noise, and it can be further mitigated by incorporating independent constraints from stellar and neutral gas probes.

Compared to the full-shape analysis of LSS (e.g., \cite{ishak2024modified}), $F_G$ provides a gravity test solely relying on the equivalence principle, and it is independent of the modeling of matter clustering and the parameterization of gravity theory. 
With $10^5$ localized FRBs and a conservative cut $k\leq 0.1\, h/{\rm Mpc}$, the overall $2\%$ accuracy of $G_{\rm light}$ measurement is already competitive to the $\Sigma_0$ constraint through full-shape modeling using all primary probes in \textit{Euclid} emission with a baseline cut $0.25\,h/{\rm Mpc}$ \cite{albuquerque2025euclid}. With more FRBs, our measurement can be pushed to the $1\%$ overall accuracy bounded by the gastrophysical systematics. Such high accuracy would not only allow us to probe the possible temporal evolution in $G_{\rm light}$, but also its spatial dependence, which encodes more information about MG including its screening effect. In summary, the proposed $F_G$ method paves the way for high precision gravity test with FRBs combining galaxy surveys, largely immune to uncertainties in modeling MG, LSS and gastrophysics. 

Throughout cosmology, the possibility to measure $\delta_m$ accurately is much more valuable than just measuring $G_{\rm light}$. Among the 4 major LSS variables ($\Psi$, $\Phi$, $\delta_m$ and $\theta_m\equiv \nabla\cdot{\bf v}_m$), if only two of them are available in observations, there would exist severe degeneracies between MG and clustered DE models \cite{kunz2007dark}. To break such degeneracies, at least three of them are demanded, although it is highly non-trivial to realize. 
In principle, $\Psi+\Phi$ can be directly measured from weak lensing, and $\Psi$ can be constructed given the measurement of peculiar velocity over multiple cosmic epochs. The velocity measurement is highly challenging currently, while it is being achieved progressively by various surveys, through the galaxy scaling relations, type Ia supernovae flux fluctuations, and luminosity distance fluctuations in bright standard sirens \cite{shi2024momentum, rosselli2025forecast, hui2006correlated}. 
However, the unbiased measurement of $\delta_m$ without assuming GR is beyond the scope of most LSS tracers including galaxy redshift-space clustering and weak lensing. Now as shown in this work, this otherwise difficult measurement can be achieved by FRB DM and the relation Eq.~(\ref{equ:be_from_bb}). With three observables (i.e., $\Psi+\Phi$, $\Psi$,  and $\delta_m$), we can construct two consistency relations that a MG model must satisfy, provided that there exists any DE model to mimic it \cite{jain2008observational}. In general, a specific MG model would either fail such tests or require fine-tuning, leading to an unambiguous distinction between the MG and DE scenarios.

\begin{acknowledgments}
This work is supported by the National Key R\&D Program of China (2023YFA1607800, 2023YFA1607801), and the Fundamental Research Funds for the Central Universities. 
This work was supported by the National Center for High-Level Talent Training in Mathematics, Physics, Chemistry, and Biology. 
This work made use of the Gravity Supercomputer at the Department of Astronomy, Shanghai Jiao Tong University. 

\end{acknowledgments}


\bibliographystyle{apsrev4-2}
\bibliography{citations}

\begin{thebibliography}{107}%
\makeatletter
\providecommand \@ifxundefined [1]{%
 \@ifx{#1\undefined}
}%
\providecommand \@ifnum [1]{%
 \ifnum #1\expandafter \@firstoftwo
 \else \expandafter \@secondoftwo
 \fi
}%
\providecommand \@ifx [1]{%
 \ifx #1\expandafter \@firstoftwo
 \else \expandafter \@secondoftwo
 \fi
}%
\providecommand \natexlab [1]{#1}%
\providecommand \enquote  [1]{``#1''}%
\providecommand \bibnamefont  [1]{#1}%
\providecommand \bibfnamefont [1]{#1}%
\providecommand \citenamefont [1]{#1}%
\providecommand \href@noop [0]{\@secondoftwo}%
\providecommand \href [0]{\begingroup \@sanitize@url \@href}%
\providecommand \@href[1]{\@@startlink{#1}\@@href}%
\providecommand \@@href[1]{\endgroup#1\@@endlink}%
\providecommand \@sanitize@url [0]{\catcode `\\12\catcode `\$12\catcode `\&12\catcode `\#12\catcode `\^12\catcode `\_12\catcode `\%12\relax}%
\providecommand \@@startlink[1]{}%
\providecommand \@@endlink[0]{}%
\providecommand \url  [0]{\begingroup\@sanitize@url \@url }%
\providecommand \@url [1]{\endgroup\@href {#1}{\urlprefix }}%
\providecommand \urlprefix  [0]{URL }%
\providecommand \Eprint [0]{\href }%
\providecommand \doibase [0]{https://doi.org/}%
\providecommand \selectlanguage [0]{\@gobble}%
\providecommand \bibinfo  [0]{\@secondoftwo}%
\providecommand \bibfield  [0]{\@secondoftwo}%
\providecommand \translation [1]{[#1]}%
\providecommand \BibitemOpen [0]{}%
\providecommand \bibitemStop [0]{}%
\providecommand \bibitemNoStop [0]{.\EOS\space}%
\providecommand \EOS [0]{\spacefactor3000\relax}%
\providecommand \BibitemShut  [1]{\csname bibitem#1\endcsname}%
\let\auto@bib@innerbib\@empty
\bibitem [{\citenamefont {Weinberg}\ \emph {et~al.}(2013)\citenamefont {Weinberg}, \citenamefont {Mortonson}, \citenamefont {Eisenstein}, \citenamefont {Hirata}, \citenamefont {Riess},\ and\ \citenamefont {Rozo}}]{weinberg2013observational}%
  \BibitemOpen
  \bibfield  {author} {\bibinfo {author} {\bibfnamefont {D.~H.}\ \bibnamefont {Weinberg}}, \bibinfo {author} {\bibfnamefont {M.~J.}\ \bibnamefont {Mortonson}}, \bibinfo {author} {\bibfnamefont {D.~J.}\ \bibnamefont {Eisenstein}}, \bibinfo {author} {\bibfnamefont {C.}~\bibnamefont {Hirata}}, \bibinfo {author} {\bibfnamefont {A.~G.}\ \bibnamefont {Riess}},\ and\ \bibinfo {author} {\bibfnamefont {E.}~\bibnamefont {Rozo}},\ }\href@noop {} {\bibfield  {journal} {\bibinfo  {journal} {Physics reports}\ }\textbf {\bibinfo {volume} {530}},\ \bibinfo {pages} {87} (\bibinfo {year} {2013})}\BibitemShut {NoStop}%
\bibitem [{\citenamefont {Kamionkowski}\ and\ \citenamefont {Riess}(2023)}]{kamionkowski2023hubble}%
  \BibitemOpen
  \bibfield  {author} {\bibinfo {author} {\bibfnamefont {M.}~\bibnamefont {Kamionkowski}}\ and\ \bibinfo {author} {\bibfnamefont {A.~G.}\ \bibnamefont {Riess}},\ }\href@noop {} {\bibfield  {journal} {\bibinfo  {journal} {Annual Review of Nuclear and Particle Science}\ }\textbf {\bibinfo {volume} {73}},\ \bibinfo {pages} {153} (\bibinfo {year} {2023})}\BibitemShut {NoStop}%
\bibitem [{\citenamefont {Jain}\ and\ \citenamefont {Zhang}(2008)}]{jain2008observational}%
  \BibitemOpen
  \bibfield  {author} {\bibinfo {author} {\bibfnamefont {B.}~\bibnamefont {Jain}}\ and\ \bibinfo {author} {\bibfnamefont {P.}~\bibnamefont {Zhang}},\ }\href@noop {} {\bibfield  {journal} {\bibinfo  {journal} {Physical Review D—Particles, Fields, Gravitation, and Cosmology}\ }\textbf {\bibinfo {volume} {78}},\ \bibinfo {pages} {063503} (\bibinfo {year} {2008})}\BibitemShut {NoStop}%
\bibitem [{\citenamefont {Clifton}\ \emph {et~al.}(2012)\citenamefont {Clifton}, \citenamefont {Ferreira}, \citenamefont {Padilla},\ and\ \citenamefont {Skordis}}]{clifton2012modified}%
  \BibitemOpen
  \bibfield  {author} {\bibinfo {author} {\bibfnamefont {T.}~\bibnamefont {Clifton}}, \bibinfo {author} {\bibfnamefont {P.~G.}\ \bibnamefont {Ferreira}}, \bibinfo {author} {\bibfnamefont {A.}~\bibnamefont {Padilla}},\ and\ \bibinfo {author} {\bibfnamefont {C.}~\bibnamefont {Skordis}},\ }\href@noop {} {\bibfield  {journal} {\bibinfo  {journal} {Physics reports}\ }\textbf {\bibinfo {volume} {513}},\ \bibinfo {pages} {1} (\bibinfo {year} {2012})}\BibitemShut {NoStop}%
\bibitem [{\citenamefont {Ezquiaga}\ and\ \citenamefont {Zumalac{\'a}rregui}(2018)}]{ezquiaga2018dark}%
  \BibitemOpen
  \bibfield  {author} {\bibinfo {author} {\bibfnamefont {J.~M.}\ \bibnamefont {Ezquiaga}}\ and\ \bibinfo {author} {\bibfnamefont {M.}~\bibnamefont {Zumalac{\'a}rregui}},\ }\href@noop {} {\bibfield  {journal} {\bibinfo  {journal} {Frontiers in Astronomy and Space Sciences}\ }\textbf {\bibinfo {volume} {5}},\ \bibinfo {pages} {44} (\bibinfo {year} {2018})}\BibitemShut {NoStop}%
\bibitem [{\citenamefont {Joyce}\ \emph {et~al.}(2015)\citenamefont {Joyce}, \citenamefont {Jain}, \citenamefont {Khoury},\ and\ \citenamefont {Trodden}}]{joyce2015beyond}%
  \BibitemOpen
  \bibfield  {author} {\bibinfo {author} {\bibfnamefont {A.}~\bibnamefont {Joyce}}, \bibinfo {author} {\bibfnamefont {B.}~\bibnamefont {Jain}}, \bibinfo {author} {\bibfnamefont {J.}~\bibnamefont {Khoury}},\ and\ \bibinfo {author} {\bibfnamefont {M.}~\bibnamefont {Trodden}},\ }\href@noop {} {\bibfield  {journal} {\bibinfo  {journal} {Physics Reports}\ }\textbf {\bibinfo {volume} {568}},\ \bibinfo {pages} {1} (\bibinfo {year} {2015})}\BibitemShut {NoStop}%
\bibitem [{\citenamefont {Koyama}(2016)}]{koyama2016cosmological}%
  \BibitemOpen
  \bibfield  {author} {\bibinfo {author} {\bibfnamefont {K.}~\bibnamefont {Koyama}},\ }\href@noop {} {\bibfield  {journal} {\bibinfo  {journal} {Reports on Progress in Physics}\ }\textbf {\bibinfo {volume} {79}},\ \bibinfo {pages} {046902} (\bibinfo {year} {2016})}\BibitemShut {NoStop}%
\bibitem [{\citenamefont {Ishak}(2019)}]{ishak2019testing}%
  \BibitemOpen
  \bibfield  {author} {\bibinfo {author} {\bibfnamefont {M.}~\bibnamefont {Ishak}},\ }\href@noop {} {\bibfield  {journal} {\bibinfo  {journal} {Living Reviews in Relativity}\ }\textbf {\bibinfo {volume} {22}},\ \bibinfo {pages} {1} (\bibinfo {year} {2019})}\BibitemShut {NoStop}%
\bibitem [{\citenamefont {Ferreira}(2019)}]{ferreira2019cosmological}%
  \BibitemOpen
  \bibfield  {author} {\bibinfo {author} {\bibfnamefont {P.~G.}\ \bibnamefont {Ferreira}},\ }\href@noop {} {\bibfield  {journal} {\bibinfo  {journal} {Annual Review of Astronomy and Astrophysics}\ }\textbf {\bibinfo {volume} {57}},\ \bibinfo {pages} {335} (\bibinfo {year} {2019})}\BibitemShut {NoStop}%
\bibitem [{\citenamefont {Karim}\ \emph {et~al.}(2025)\citenamefont {Karim}, \citenamefont {Aguilar}, \citenamefont {Ahlen}, \citenamefont {Alam}, \citenamefont {Allen}, \citenamefont {Allende~Prieto}, \citenamefont {Alves}, \citenamefont {Anand}, \citenamefont {Andrade}, \citenamefont {Armengaud} \emph {et~al.}}]{karim2025desi}%
  \BibitemOpen
  \bibfield  {author} {\bibinfo {author} {\bibfnamefont {M.~A.}\ \bibnamefont {Karim}}, \bibinfo {author} {\bibfnamefont {J.}~\bibnamefont {Aguilar}}, \bibinfo {author} {\bibfnamefont {S.}~\bibnamefont {Ahlen}}, \bibinfo {author} {\bibfnamefont {S.}~\bibnamefont {Alam}}, \bibinfo {author} {\bibfnamefont {L.}~\bibnamefont {Allen}}, \bibinfo {author} {\bibfnamefont {C.}~\bibnamefont {Allende~Prieto}}, \bibinfo {author} {\bibfnamefont {O.}~\bibnamefont {Alves}}, \bibinfo {author} {\bibfnamefont {A.}~\bibnamefont {Anand}}, \bibinfo {author} {\bibfnamefont {U.}~\bibnamefont {Andrade}}, \bibinfo {author} {\bibfnamefont {E.}~\bibnamefont {Armengaud}}, \emph {et~al.},\ }\href@noop {} {\bibfield  {journal} {\bibinfo  {journal} {arXiv e-prints}\ ,\ \bibinfo {pages} {arXiv}} (\bibinfo {year} {2025})}\BibitemShut {NoStop}%
\bibitem [{\citenamefont {Lodha}\ \emph {et~al.}(2025)\citenamefont {Lodha}, \citenamefont {Calderon}, \citenamefont {Matthewson}, \citenamefont {Shafieloo}, \citenamefont {Ishak}, \citenamefont {Pan}, \citenamefont {Garcia-Quintero}, \citenamefont {Huterer}, \citenamefont {Valogiannis}, \citenamefont {Ure{\~n}a-L{\'o}pez} \emph {et~al.}}]{lodha2025extended}%
  \BibitemOpen
  \bibfield  {author} {\bibinfo {author} {\bibfnamefont {K.}~\bibnamefont {Lodha}}, \bibinfo {author} {\bibfnamefont {R.}~\bibnamefont {Calderon}}, \bibinfo {author} {\bibfnamefont {W.}~\bibnamefont {Matthewson}}, \bibinfo {author} {\bibfnamefont {A.}~\bibnamefont {Shafieloo}}, \bibinfo {author} {\bibfnamefont {M.}~\bibnamefont {Ishak}}, \bibinfo {author} {\bibfnamefont {J.}~\bibnamefont {Pan}}, \bibinfo {author} {\bibfnamefont {C.}~\bibnamefont {Garcia-Quintero}}, \bibinfo {author} {\bibfnamefont {D.}~\bibnamefont {Huterer}}, \bibinfo {author} {\bibfnamefont {G.}~\bibnamefont {Valogiannis}}, \bibinfo {author} {\bibfnamefont {L.}~\bibnamefont {Ure{\~n}a-L{\'o}pez}}, \emph {et~al.},\ }\href@noop {} {\bibfield  {journal} {\bibinfo  {journal} {arXiv preprint arXiv:2503.14743}\ } (\bibinfo {year} {2025})}\BibitemShut {NoStop}%
\bibitem [{\citenamefont {Gu}\ \emph {et~al.}(2025)\citenamefont {Gu}, \citenamefont {Wang}, \citenamefont {Wang}, \citenamefont {Zhao}, \citenamefont {Pogosian}, \citenamefont {Koyama}, \citenamefont {Peacock}, \citenamefont {Cai}, \citenamefont {Cervantes-Cot}, \citenamefont {Ishak} \emph {et~al.}}]{gu2025dynamical}%
  \BibitemOpen
  \bibfield  {author} {\bibinfo {author} {\bibfnamefont {G.}~\bibnamefont {Gu}}, \bibinfo {author} {\bibfnamefont {X.}~\bibnamefont {Wang}}, \bibinfo {author} {\bibfnamefont {Y.}~\bibnamefont {Wang}}, \bibinfo {author} {\bibfnamefont {G.-B.}\ \bibnamefont {Zhao}}, \bibinfo {author} {\bibfnamefont {L.}~\bibnamefont {Pogosian}}, \bibinfo {author} {\bibfnamefont {K.}~\bibnamefont {Koyama}}, \bibinfo {author} {\bibfnamefont {J.~A.}\ \bibnamefont {Peacock}}, \bibinfo {author} {\bibfnamefont {Z.}~\bibnamefont {Cai}}, \bibinfo {author} {\bibfnamefont {J.~L.}\ \bibnamefont {Cervantes-Cot}}, \bibinfo {author} {\bibfnamefont {M.}~\bibnamefont {Ishak}}, \emph {et~al.},\ }\href@noop {} {\bibfield  {journal} {\bibinfo  {journal} {Nature Astronomy}\ ,\ \bibinfo {pages} {1}} (\bibinfo {year} {2025})}\BibitemShut {NoStop}%
\bibitem [{\citenamefont {Zhang}\ \emph {et~al.}(2007)\citenamefont {Zhang}, \citenamefont {Liguori}, \citenamefont {Bean},\ and\ \citenamefont {Dodelson}}]{zhang2007probing}%
  \BibitemOpen
  \bibfield  {author} {\bibinfo {author} {\bibfnamefont {P.}~\bibnamefont {Zhang}}, \bibinfo {author} {\bibfnamefont {M.}~\bibnamefont {Liguori}}, \bibinfo {author} {\bibfnamefont {R.}~\bibnamefont {Bean}},\ and\ \bibinfo {author} {\bibfnamefont {S.}~\bibnamefont {Dodelson}},\ }\href@noop {} {\bibfield  {journal} {\bibinfo  {journal} {Physical Review Letters}\ }\textbf {\bibinfo {volume} {99}},\ \bibinfo {pages} {141302} (\bibinfo {year} {2007})}\BibitemShut {NoStop}%
\bibitem [{\citenamefont {Zhao}\ \emph {et~al.}(2009)\citenamefont {Zhao}, \citenamefont {Pogosian}, \citenamefont {Silvestri},\ and\ \citenamefont {Zylberberg}}]{zhao2009searching}%
  \BibitemOpen
  \bibfield  {author} {\bibinfo {author} {\bibfnamefont {G.-B.}\ \bibnamefont {Zhao}}, \bibinfo {author} {\bibfnamefont {L.}~\bibnamefont {Pogosian}}, \bibinfo {author} {\bibfnamefont {A.}~\bibnamefont {Silvestri}},\ and\ \bibinfo {author} {\bibfnamefont {J.}~\bibnamefont {Zylberberg}},\ }\href@noop {} {\bibfield  {journal} {\bibinfo  {journal} {Physical Review D—Particles, Fields, Gravitation, and Cosmology}\ }\textbf {\bibinfo {volume} {79}},\ \bibinfo {pages} {083513} (\bibinfo {year} {2009})}\BibitemShut {NoStop}%
\bibitem [{\citenamefont {Levon}\ and\ \citenamefont {Alessandra}(2016)}]{levon2016can}%
  \BibitemOpen
  \bibfield  {author} {\bibinfo {author} {\bibfnamefont {P.}~\bibnamefont {Levon}}\ and\ \bibinfo {author} {\bibfnamefont {S.}~\bibnamefont {Alessandra}},\ }\href@noop {} {\bibfield  {journal} {\bibinfo  {journal} {Phys. Rev. D}\ }\textbf {\bibinfo {volume} {94}},\ \bibinfo {pages} {104014} (\bibinfo {year} {2016})}\BibitemShut {NoStop}%
\bibitem [{\citenamefont {Ade}\ \emph {et~al.}(2016)\citenamefont {Ade}, \citenamefont {Aghanim}, \citenamefont {Arnaud}, \citenamefont {Ashdown}, \citenamefont {Aumont}, \citenamefont {Baccigalupi}, \citenamefont {Banday}, \citenamefont {Barreiro}, \citenamefont {Bartolo}, \citenamefont {Battaner} \emph {et~al.}}]{ade2016planck}%
  \BibitemOpen
  \bibfield  {author} {\bibinfo {author} {\bibfnamefont {P.~A.}\ \bibnamefont {Ade}}, \bibinfo {author} {\bibfnamefont {N.}~\bibnamefont {Aghanim}}, \bibinfo {author} {\bibfnamefont {M.}~\bibnamefont {Arnaud}}, \bibinfo {author} {\bibfnamefont {M.}~\bibnamefont {Ashdown}}, \bibinfo {author} {\bibfnamefont {J.}~\bibnamefont {Aumont}}, \bibinfo {author} {\bibfnamefont {C.}~\bibnamefont {Baccigalupi}}, \bibinfo {author} {\bibfnamefont {A.}~\bibnamefont {Banday}}, \bibinfo {author} {\bibfnamefont {R.}~\bibnamefont {Barreiro}}, \bibinfo {author} {\bibfnamefont {N.}~\bibnamefont {Bartolo}}, \bibinfo {author} {\bibfnamefont {E.}~\bibnamefont {Battaner}}, \emph {et~al.},\ }\href@noop {} {\bibfield  {journal} {\bibinfo  {journal} {Astronomy \& Astrophysics}\ }\textbf {\bibinfo {volume} {594}},\ \bibinfo {pages} {A14} (\bibinfo {year} {2016})}\BibitemShut {NoStop}%
\bibitem [{\citenamefont {Alam}\ \emph {et~al.}(2021)\citenamefont {Alam}, \citenamefont {Aubert}, \citenamefont {Avila}, \citenamefont {Balland}, \citenamefont {Bautista}, \citenamefont {Bershady}, \citenamefont {Bizyaev}, \citenamefont {Blanton}, \citenamefont {Bolton}, \citenamefont {Bovy} \emph {et~al.}}]{alam2021completed}%
  \BibitemOpen
  \bibfield  {author} {\bibinfo {author} {\bibfnamefont {S.}~\bibnamefont {Alam}}, \bibinfo {author} {\bibfnamefont {M.}~\bibnamefont {Aubert}}, \bibinfo {author} {\bibfnamefont {S.}~\bibnamefont {Avila}}, \bibinfo {author} {\bibfnamefont {C.}~\bibnamefont {Balland}}, \bibinfo {author} {\bibfnamefont {J.~E.}\ \bibnamefont {Bautista}}, \bibinfo {author} {\bibfnamefont {M.~A.}\ \bibnamefont {Bershady}}, \bibinfo {author} {\bibfnamefont {D.}~\bibnamefont {Bizyaev}}, \bibinfo {author} {\bibfnamefont {M.~R.}\ \bibnamefont {Blanton}}, \bibinfo {author} {\bibfnamefont {A.~S.}\ \bibnamefont {Bolton}}, \bibinfo {author} {\bibfnamefont {J.}~\bibnamefont {Bovy}}, \emph {et~al.},\ }\href@noop {} {\bibfield  {journal} {\bibinfo  {journal} {Physical Review D}\ }\textbf {\bibinfo {volume} {103}},\ \bibinfo {pages} {083533} (\bibinfo {year} {2021})}\BibitemShut {NoStop}%
\bibitem [{\citenamefont {Ishak}\ \emph {et~al.}(2024)\citenamefont {Ishak}, \citenamefont {Pan}, \citenamefont {Calderon}, \citenamefont {Lodha}, \citenamefont {Valogiannis}, \citenamefont {Aviles}, \citenamefont {Niz}, \citenamefont {Yi}, \citenamefont {Zheng}, \citenamefont {Garcia-Quintero} \emph {et~al.}}]{ishak2024modified}%
  \BibitemOpen
  \bibfield  {author} {\bibinfo {author} {\bibfnamefont {M.}~\bibnamefont {Ishak}}, \bibinfo {author} {\bibfnamefont {J.}~\bibnamefont {Pan}}, \bibinfo {author} {\bibfnamefont {R.}~\bibnamefont {Calderon}}, \bibinfo {author} {\bibfnamefont {K.}~\bibnamefont {Lodha}}, \bibinfo {author} {\bibfnamefont {G.}~\bibnamefont {Valogiannis}}, \bibinfo {author} {\bibfnamefont {A.}~\bibnamefont {Aviles}}, \bibinfo {author} {\bibfnamefont {G.}~\bibnamefont {Niz}}, \bibinfo {author} {\bibfnamefont {L.}~\bibnamefont {Yi}}, \bibinfo {author} {\bibfnamefont {C.}~\bibnamefont {Zheng}}, \bibinfo {author} {\bibfnamefont {C.}~\bibnamefont {Garcia-Quintero}}, \emph {et~al.},\ }\href@noop {} {\bibfield  {journal} {\bibinfo  {journal} {arXiv preprint arXiv:2411.12026}\ } (\bibinfo {year} {2024})}\BibitemShut {NoStop}%
\bibitem [{\citenamefont {Reyes}\ \emph {et~al.}(2010)\citenamefont {Reyes}, \citenamefont {Mandelbaum}, \citenamefont {Seljak}, \citenamefont {Baldauf}, \citenamefont {Gunn}, \citenamefont {Lombriser},\ and\ \citenamefont {Smith}}]{reyes2010confirmation}%
  \BibitemOpen
  \bibfield  {author} {\bibinfo {author} {\bibfnamefont {R.}~\bibnamefont {Reyes}}, \bibinfo {author} {\bibfnamefont {R.}~\bibnamefont {Mandelbaum}}, \bibinfo {author} {\bibfnamefont {U.}~\bibnamefont {Seljak}}, \bibinfo {author} {\bibfnamefont {T.}~\bibnamefont {Baldauf}}, \bibinfo {author} {\bibfnamefont {J.~E.}\ \bibnamefont {Gunn}}, \bibinfo {author} {\bibfnamefont {L.}~\bibnamefont {Lombriser}},\ and\ \bibinfo {author} {\bibfnamefont {R.~E.}\ \bibnamefont {Smith}},\ }\href@noop {} {\bibfield  {journal} {\bibinfo  {journal} {Nature}\ }\textbf {\bibinfo {volume} {464}},\ \bibinfo {pages} {256} (\bibinfo {year} {2010})}\BibitemShut {NoStop}%
\bibitem [{\citenamefont {Pullen}\ \emph {et~al.}(2016)\citenamefont {Pullen}, \citenamefont {Alam}, \citenamefont {He},\ and\ \citenamefont {Ho}}]{pullen2016constraining}%
  \BibitemOpen
  \bibfield  {author} {\bibinfo {author} {\bibfnamefont {A.~R.}\ \bibnamefont {Pullen}}, \bibinfo {author} {\bibfnamefont {S.}~\bibnamefont {Alam}}, \bibinfo {author} {\bibfnamefont {S.}~\bibnamefont {He}},\ and\ \bibinfo {author} {\bibfnamefont {S.}~\bibnamefont {Ho}},\ }\href@noop {} {\bibfield  {journal} {\bibinfo  {journal} {Monthly Notices of the Royal Astronomical Society}\ }\textbf {\bibinfo {volume} {460}},\ \bibinfo {pages} {4098} (\bibinfo {year} {2016})}\BibitemShut {NoStop}%
\bibitem [{\citenamefont {Amon}\ \emph {et~al.}(2018)\citenamefont {Amon}, \citenamefont {Blake}, \citenamefont {Heymans}, \citenamefont {Leonard}, \citenamefont {Asgari}, \citenamefont {Bilicki}, \citenamefont {Choi}, \citenamefont {Erben}, \citenamefont {Glazebrook}, \citenamefont {Harnois-Deraps} \emph {et~al.}}]{amon2018kids}%
  \BibitemOpen
  \bibfield  {author} {\bibinfo {author} {\bibfnamefont {A.}~\bibnamefont {Amon}}, \bibinfo {author} {\bibfnamefont {C.}~\bibnamefont {Blake}}, \bibinfo {author} {\bibfnamefont {C.}~\bibnamefont {Heymans}}, \bibinfo {author} {\bibfnamefont {C.}~\bibnamefont {Leonard}}, \bibinfo {author} {\bibfnamefont {M.}~\bibnamefont {Asgari}}, \bibinfo {author} {\bibfnamefont {M.}~\bibnamefont {Bilicki}}, \bibinfo {author} {\bibfnamefont {A.}~\bibnamefont {Choi}}, \bibinfo {author} {\bibfnamefont {T.}~\bibnamefont {Erben}}, \bibinfo {author} {\bibfnamefont {K.}~\bibnamefont {Glazebrook}}, \bibinfo {author} {\bibfnamefont {J.}~\bibnamefont {Harnois-Deraps}}, \emph {et~al.},\ }\href@noop {} {\bibfield  {journal} {\bibinfo  {journal} {Monthly Notices of the Royal Astronomical Society}\ }\textbf {\bibinfo {volume} {479}},\ \bibinfo {pages} {3422} (\bibinfo {year} {2018})}\BibitemShut {NoStop}%
\bibitem [{\citenamefont {Singh}\ \emph {et~al.}(2019)\citenamefont {Singh}, \citenamefont {Alam}, \citenamefont {Mandelbaum}, \citenamefont {Seljak}, \citenamefont {Rodriguez-Torres},\ and\ \citenamefont {Ho}}]{singh2019probing}%
  \BibitemOpen
  \bibfield  {author} {\bibinfo {author} {\bibfnamefont {S.}~\bibnamefont {Singh}}, \bibinfo {author} {\bibfnamefont {S.}~\bibnamefont {Alam}}, \bibinfo {author} {\bibfnamefont {R.}~\bibnamefont {Mandelbaum}}, \bibinfo {author} {\bibfnamefont {U.}~\bibnamefont {Seljak}}, \bibinfo {author} {\bibfnamefont {S.}~\bibnamefont {Rodriguez-Torres}},\ and\ \bibinfo {author} {\bibfnamefont {S.}~\bibnamefont {Ho}},\ }\href@noop {} {\bibfield  {journal} {\bibinfo  {journal} {Monthly Notices of the Royal Astronomical Society}\ }\textbf {\bibinfo {volume} {482}},\ \bibinfo {pages} {785} (\bibinfo {year} {2019})}\BibitemShut {NoStop}%
\bibitem [{\citenamefont {Zhang}\ \emph {et~al.}(2021)\citenamefont {Zhang}, \citenamefont {Pullen}, \citenamefont {Alam}, \citenamefont {Singh}, \citenamefont {Burtin}, \citenamefont {Chuang}, \citenamefont {Hou}, \citenamefont {Lyke}, \citenamefont {Myers}, \citenamefont {Neveux} \emph {et~al.}}]{zhang2021testing}%
  \BibitemOpen
  \bibfield  {author} {\bibinfo {author} {\bibfnamefont {Y.}~\bibnamefont {Zhang}}, \bibinfo {author} {\bibfnamefont {A.~R.}\ \bibnamefont {Pullen}}, \bibinfo {author} {\bibfnamefont {S.}~\bibnamefont {Alam}}, \bibinfo {author} {\bibfnamefont {S.}~\bibnamefont {Singh}}, \bibinfo {author} {\bibfnamefont {E.}~\bibnamefont {Burtin}}, \bibinfo {author} {\bibfnamefont {C.-H.}\ \bibnamefont {Chuang}}, \bibinfo {author} {\bibfnamefont {J.}~\bibnamefont {Hou}}, \bibinfo {author} {\bibfnamefont {B.~W.}\ \bibnamefont {Lyke}}, \bibinfo {author} {\bibfnamefont {A.~D.}\ \bibnamefont {Myers}}, \bibinfo {author} {\bibfnamefont {R.}~\bibnamefont {Neveux}}, \emph {et~al.},\ }\href@noop {} {\bibfield  {journal} {\bibinfo  {journal} {Monthly Notices of the Royal Astronomical Society}\ }\textbf {\bibinfo {volume} {501}},\ \bibinfo {pages} {1013} (\bibinfo {year} {2021})}\BibitemShut {NoStop}%
\bibitem [{\citenamefont {Wenzl}\ \emph {et~al.}(2024)\citenamefont {Wenzl}, \citenamefont {Bean}, \citenamefont {Chen}, \citenamefont {Farren}, \citenamefont {Madhavacheril}, \citenamefont {Marques}, \citenamefont {Qu}, \citenamefont {Sehgal}, \citenamefont {Sherwin},\ and\ \citenamefont {Van~Engelen}}]{wenzl2024constraining}%
  \BibitemOpen
  \bibfield  {author} {\bibinfo {author} {\bibfnamefont {L.}~\bibnamefont {Wenzl}}, \bibinfo {author} {\bibfnamefont {R.}~\bibnamefont {Bean}}, \bibinfo {author} {\bibfnamefont {S.-F.}\ \bibnamefont {Chen}}, \bibinfo {author} {\bibfnamefont {G.~S.}\ \bibnamefont {Farren}}, \bibinfo {author} {\bibfnamefont {M.~S.}\ \bibnamefont {Madhavacheril}}, \bibinfo {author} {\bibfnamefont {G.~A.}\ \bibnamefont {Marques}}, \bibinfo {author} {\bibfnamefont {F.~J.}\ \bibnamefont {Qu}}, \bibinfo {author} {\bibfnamefont {N.}~\bibnamefont {Sehgal}}, \bibinfo {author} {\bibfnamefont {B.~D.}\ \bibnamefont {Sherwin}},\ and\ \bibinfo {author} {\bibfnamefont {A.}~\bibnamefont {Van~Engelen}},\ }\href@noop {} {\bibfield  {journal} {\bibinfo  {journal} {Physical Review D}\ }\textbf {\bibinfo {volume} {109}},\ \bibinfo {pages} {083540} (\bibinfo {year} {2024})}\BibitemShut {NoStop}%
\bibitem [{\citenamefont {Wenzl}\ \emph {et~al.}(2025)\citenamefont {Wenzl}, \citenamefont {An}, \citenamefont {Battaglia}, \citenamefont {Bean}, \citenamefont {Calabrese}, \citenamefont {Chen}, \citenamefont {Choi}, \citenamefont {Darwish}, \citenamefont {Dunkley}, \citenamefont {Farren} \emph {et~al.}}]{wenzl2025atacama}%
  \BibitemOpen
  \bibfield  {author} {\bibinfo {author} {\bibfnamefont {L.}~\bibnamefont {Wenzl}}, \bibinfo {author} {\bibfnamefont {R.}~\bibnamefont {An}}, \bibinfo {author} {\bibfnamefont {N.}~\bibnamefont {Battaglia}}, \bibinfo {author} {\bibfnamefont {R.}~\bibnamefont {Bean}}, \bibinfo {author} {\bibfnamefont {E.}~\bibnamefont {Calabrese}}, \bibinfo {author} {\bibfnamefont {S.-F.}\ \bibnamefont {Chen}}, \bibinfo {author} {\bibfnamefont {S.~K.}\ \bibnamefont {Choi}}, \bibinfo {author} {\bibfnamefont {O.}~\bibnamefont {Darwish}}, \bibinfo {author} {\bibfnamefont {J.}~\bibnamefont {Dunkley}}, \bibinfo {author} {\bibfnamefont {G.~S.}\ \bibnamefont {Farren}}, \emph {et~al.},\ }\href@noop {} {\bibfield  {journal} {\bibinfo  {journal} {Physical Review D}\ }\textbf {\bibinfo {volume} {111}},\ \bibinfo {pages} {043535} (\bibinfo {year} {2025})}\BibitemShut {NoStop}%
\bibitem [{\citenamefont {Jullo}\ \emph {et~al.}(2019)\citenamefont {Jullo}, \citenamefont {De~La~Torre}, \citenamefont {Cousinou}, \citenamefont {Escoffier}, \citenamefont {Giocoli}, \citenamefont {Metcalf}, \citenamefont {Comparat}, \citenamefont {Shan}, \citenamefont {Makler}, \citenamefont {Kneib} \emph {et~al.}}]{jullo2019testing}%
  \BibitemOpen
  \bibfield  {author} {\bibinfo {author} {\bibfnamefont {E.}~\bibnamefont {Jullo}}, \bibinfo {author} {\bibfnamefont {S.}~\bibnamefont {De~La~Torre}}, \bibinfo {author} {\bibfnamefont {M.-C.}\ \bibnamefont {Cousinou}}, \bibinfo {author} {\bibfnamefont {S.}~\bibnamefont {Escoffier}}, \bibinfo {author} {\bibfnamefont {C.}~\bibnamefont {Giocoli}}, \bibinfo {author} {\bibfnamefont {R.~B.}\ \bibnamefont {Metcalf}}, \bibinfo {author} {\bibfnamefont {J.}~\bibnamefont {Comparat}}, \bibinfo {author} {\bibfnamefont {H.-Y.}\ \bibnamefont {Shan}}, \bibinfo {author} {\bibfnamefont {M.}~\bibnamefont {Makler}}, \bibinfo {author} {\bibfnamefont {J.-P.}\ \bibnamefont {Kneib}}, \emph {et~al.},\ }\href@noop {} {\bibfield  {journal} {\bibinfo  {journal} {Astronomy \& Astrophysics}\ }\textbf {\bibinfo {volume} {627}},\ \bibinfo {pages} {A137} (\bibinfo {year} {2019})}\BibitemShut {NoStop}%
\bibitem [{\citenamefont {Blake}\ \emph {et~al.}(2020)\citenamefont {Blake}, \citenamefont {Amon}, \citenamefont {Asgari}, \citenamefont {Bilicki}, \citenamefont {Dvornik}, \citenamefont {Erben}, \citenamefont {Giblin}, \citenamefont {Glazebrook}, \citenamefont {Heymans}, \citenamefont {Hildebrandt} \emph {et~al.}}]{blake2020testing}%
  \BibitemOpen
  \bibfield  {author} {\bibinfo {author} {\bibfnamefont {C.}~\bibnamefont {Blake}}, \bibinfo {author} {\bibfnamefont {A.}~\bibnamefont {Amon}}, \bibinfo {author} {\bibfnamefont {M.}~\bibnamefont {Asgari}}, \bibinfo {author} {\bibfnamefont {M.}~\bibnamefont {Bilicki}}, \bibinfo {author} {\bibfnamefont {A.}~\bibnamefont {Dvornik}}, \bibinfo {author} {\bibfnamefont {T.}~\bibnamefont {Erben}}, \bibinfo {author} {\bibfnamefont {B.}~\bibnamefont {Giblin}}, \bibinfo {author} {\bibfnamefont {K.}~\bibnamefont {Glazebrook}}, \bibinfo {author} {\bibfnamefont {C.}~\bibnamefont {Heymans}}, \bibinfo {author} {\bibfnamefont {H.}~\bibnamefont {Hildebrandt}}, \emph {et~al.},\ }\href@noop {} {\bibfield  {journal} {\bibinfo  {journal} {Astronomy \& Astrophysics}\ }\textbf {\bibinfo {volume} {642}},\ \bibinfo {pages} {A158} (\bibinfo {year} {2020})}\BibitemShut {NoStop}%
\bibitem [{\citenamefont {Alam}\ \emph {et~al.}(2017)\citenamefont {Alam}, \citenamefont {Miyatake}, \citenamefont {More}, \citenamefont {Ho},\ and\ \citenamefont {Mandelbaum}}]{alam2017testing}%
  \BibitemOpen
  \bibfield  {author} {\bibinfo {author} {\bibfnamefont {S.}~\bibnamefont {Alam}}, \bibinfo {author} {\bibfnamefont {H.}~\bibnamefont {Miyatake}}, \bibinfo {author} {\bibfnamefont {S.}~\bibnamefont {More}}, \bibinfo {author} {\bibfnamefont {S.}~\bibnamefont {Ho}},\ and\ \bibinfo {author} {\bibfnamefont {R.}~\bibnamefont {Mandelbaum}},\ }\href@noop {} {\bibfield  {journal} {\bibinfo  {journal} {Monthly Notices of the Royal Astronomical Society}\ }\textbf {\bibinfo {volume} {465}},\ \bibinfo {pages} {4853} (\bibinfo {year} {2017})}\BibitemShut {NoStop}%
\bibitem [{\citenamefont {Blake}\ \emph {et~al.}(2016)\citenamefont {Blake}, \citenamefont {Joudaki}, \citenamefont {Heymans}, \citenamefont {Choi}, \citenamefont {Erben}, \citenamefont {Harnois-Deraps}, \citenamefont {Hildebrandt}, \citenamefont {Joachimi}, \citenamefont {Nakajima}, \citenamefont {van Waerbeke} \emph {et~al.}}]{blake2016rcslens}%
  \BibitemOpen
  \bibfield  {author} {\bibinfo {author} {\bibfnamefont {C.}~\bibnamefont {Blake}}, \bibinfo {author} {\bibfnamefont {S.}~\bibnamefont {Joudaki}}, \bibinfo {author} {\bibfnamefont {C.}~\bibnamefont {Heymans}}, \bibinfo {author} {\bibfnamefont {A.}~\bibnamefont {Choi}}, \bibinfo {author} {\bibfnamefont {T.}~\bibnamefont {Erben}}, \bibinfo {author} {\bibfnamefont {J.}~\bibnamefont {Harnois-Deraps}}, \bibinfo {author} {\bibfnamefont {H.}~\bibnamefont {Hildebrandt}}, \bibinfo {author} {\bibfnamefont {B.}~\bibnamefont {Joachimi}}, \bibinfo {author} {\bibfnamefont {R.}~\bibnamefont {Nakajima}}, \bibinfo {author} {\bibfnamefont {L.}~\bibnamefont {van Waerbeke}}, \emph {et~al.},\ }\href@noop {} {\bibfield  {journal} {\bibinfo  {journal} {Monthly Notices of the Royal Astronomical Society}\ }\textbf {\bibinfo {volume} {456}},\ \bibinfo {pages} {2806} (\bibinfo {year} {2016})}\BibitemShut {NoStop}%
\bibitem [{\citenamefont {Rauhut}\ \emph {et~al.}(2025)\citenamefont {Rauhut}, \citenamefont {Blake}, \citenamefont {Andrade}, \citenamefont {Noriega}, \citenamefont {Aguilar}, \citenamefont {Ahlen}, \citenamefont {BenZvi}, \citenamefont {Bianchi}, \citenamefont {Brooks}, \citenamefont {Claybaugh} \emph {et~al.}}]{rauhut2025testing}%
  \BibitemOpen
  \bibfield  {author} {\bibinfo {author} {\bibfnamefont {S.}~\bibnamefont {Rauhut}}, \bibinfo {author} {\bibfnamefont {C.}~\bibnamefont {Blake}}, \bibinfo {author} {\bibfnamefont {U.}~\bibnamefont {Andrade}}, \bibinfo {author} {\bibfnamefont {H.}~\bibnamefont {Noriega}}, \bibinfo {author} {\bibfnamefont {J.}~\bibnamefont {Aguilar}}, \bibinfo {author} {\bibfnamefont {S.}~\bibnamefont {Ahlen}}, \bibinfo {author} {\bibfnamefont {S.}~\bibnamefont {BenZvi}}, \bibinfo {author} {\bibfnamefont {D.}~\bibnamefont {Bianchi}}, \bibinfo {author} {\bibfnamefont {D.}~\bibnamefont {Brooks}}, \bibinfo {author} {\bibfnamefont {T.}~\bibnamefont {Claybaugh}}, \emph {et~al.},\ }\href@noop {} {\bibfield  {journal} {\bibinfo  {journal} {arXiv preprint arXiv:2507.16098}\ } (\bibinfo {year} {2025})}\BibitemShut {NoStop}%
\bibitem [{\citenamefont {Li}\ and\ \citenamefont {Xia}(2025)}]{li2025testing}%
  \BibitemOpen
  \bibfield  {author} {\bibinfo {author} {\bibfnamefont {S.}~\bibnamefont {Li}}\ and\ \bibinfo {author} {\bibfnamefont {J.-Q.}\ \bibnamefont {Xia}},\ }\href@noop {} {\bibfield  {journal} {\bibinfo  {journal} {The Astrophysical Journal Supplement Series}\ }\textbf {\bibinfo {volume} {276}},\ \bibinfo {pages} {71} (\bibinfo {year} {2025})}\BibitemShut {NoStop}%
\bibitem [{\citenamefont {Ioka}(2003)}]{ioka2003cosmic}%
  \BibitemOpen
  \bibfield  {author} {\bibinfo {author} {\bibfnamefont {K.}~\bibnamefont {Ioka}},\ }\href@noop {} {\bibfield  {journal} {\bibinfo  {journal} {The Astrophysical Journal}\ }\textbf {\bibinfo {volume} {598}},\ \bibinfo {pages} {L79} (\bibinfo {year} {2003})}\BibitemShut {NoStop}%
\bibitem [{\citenamefont {Zhang}(2023)}]{zhang2023physics}%
  \BibitemOpen
  \bibfield  {author} {\bibinfo {author} {\bibfnamefont {B.}~\bibnamefont {Zhang}},\ }\href@noop {} {\bibfield  {journal} {\bibinfo  {journal} {Reviews of Modern Physics}\ }\textbf {\bibinfo {volume} {95}},\ \bibinfo {pages} {035005} (\bibinfo {year} {2023})}\BibitemShut {NoStop}%
\bibitem [{\citenamefont {Macquart}\ \emph {et~al.}(2020)\citenamefont {Macquart}, \citenamefont {Prochaska}, \citenamefont {McQuinn}, \citenamefont {Bannister}, \citenamefont {Bhandari}, \citenamefont {Day}, \citenamefont {Deller}, \citenamefont {Ekers}, \citenamefont {James}, \citenamefont {Marnoch} \emph {et~al.}}]{macquart2020census}%
  \BibitemOpen
  \bibfield  {author} {\bibinfo {author} {\bibfnamefont {J.-P.}\ \bibnamefont {Macquart}}, \bibinfo {author} {\bibfnamefont {J.}~\bibnamefont {Prochaska}}, \bibinfo {author} {\bibfnamefont {M.}~\bibnamefont {McQuinn}}, \bibinfo {author} {\bibfnamefont {K.}~\bibnamefont {Bannister}}, \bibinfo {author} {\bibfnamefont {S.}~\bibnamefont {Bhandari}}, \bibinfo {author} {\bibfnamefont {C.}~\bibnamefont {Day}}, \bibinfo {author} {\bibfnamefont {A.}~\bibnamefont {Deller}}, \bibinfo {author} {\bibfnamefont {R.}~\bibnamefont {Ekers}}, \bibinfo {author} {\bibfnamefont {C.}~\bibnamefont {James}}, \bibinfo {author} {\bibfnamefont {L.}~\bibnamefont {Marnoch}}, \emph {et~al.},\ }\href@noop {} {\bibfield  {journal} {\bibinfo  {journal} {Nature}\ }\textbf {\bibinfo {volume} {581}},\ \bibinfo {pages} {391} (\bibinfo {year} {2020})}\BibitemShut {NoStop}%
\bibitem [{\citenamefont {Connor}\ \emph {et~al.}(2025)\citenamefont {Connor}, \citenamefont {Ravi}, \citenamefont {Sharma}, \citenamefont {Ocker}, \citenamefont {Faber}, \citenamefont {Hallinan}, \citenamefont {Harnach}, \citenamefont {Hellbourg}, \citenamefont {Hobbs}, \citenamefont {Hodge} \emph {et~al.}}]{connor2025gas}%
  \BibitemOpen
  \bibfield  {author} {\bibinfo {author} {\bibfnamefont {L.}~\bibnamefont {Connor}}, \bibinfo {author} {\bibfnamefont {V.}~\bibnamefont {Ravi}}, \bibinfo {author} {\bibfnamefont {K.}~\bibnamefont {Sharma}}, \bibinfo {author} {\bibfnamefont {S.~K.}\ \bibnamefont {Ocker}}, \bibinfo {author} {\bibfnamefont {J.}~\bibnamefont {Faber}}, \bibinfo {author} {\bibfnamefont {G.}~\bibnamefont {Hallinan}}, \bibinfo {author} {\bibfnamefont {C.}~\bibnamefont {Harnach}}, \bibinfo {author} {\bibfnamefont {G.}~\bibnamefont {Hellbourg}}, \bibinfo {author} {\bibfnamefont {R.}~\bibnamefont {Hobbs}}, \bibinfo {author} {\bibfnamefont {D.}~\bibnamefont {Hodge}}, \emph {et~al.},\ }\href@noop {} {\bibfield  {journal} {\bibinfo  {journal} {Nature Astronomy}\ ,\ \bibinfo {pages} {1}} (\bibinfo {year} {2025})}\BibitemShut {NoStop}%
\bibitem [{\citenamefont {Pen}(2018)}]{pen2018nature}%
  \BibitemOpen
  \bibfield  {author} {\bibinfo {author} {\bibfnamefont {U.-L.}\ \bibnamefont {Pen}},\ }\href@noop {} {\bibfield  {journal} {\bibinfo  {journal} {Nature Astronomy}\ }\textbf {\bibinfo {volume} {2}},\ \bibinfo {pages} {842} (\bibinfo {year} {2018})}\BibitemShut {NoStop}%
\bibitem [{\citenamefont {Fialkov}\ and\ \citenamefont {Loeb}(2017)}]{fialkov2017fast}%
  \BibitemOpen
  \bibfield  {author} {\bibinfo {author} {\bibfnamefont {A.}~\bibnamefont {Fialkov}}\ and\ \bibinfo {author} {\bibfnamefont {A.}~\bibnamefont {Loeb}},\ }\href@noop {} {\bibfield  {journal} {\bibinfo  {journal} {The Astrophysical Journal Letters}\ }\textbf {\bibinfo {volume} {846}},\ \bibinfo {pages} {L27} (\bibinfo {year} {2017})}\BibitemShut {NoStop}%
\bibitem [{\citenamefont {Hallinan}\ \emph {et~al.}(2019)\citenamefont {Hallinan}, \citenamefont {Ravi}, \citenamefont {Weinreb}, \citenamefont {Kocz}, \citenamefont {Huang}, \citenamefont {Woody}, \citenamefont {Lamb}, \citenamefont {D’Addario}, \citenamefont {Catha}, \citenamefont {Shi} \emph {et~al.}}]{hallinan2019astro2020}%
  \BibitemOpen
  \bibfield  {author} {\bibinfo {author} {\bibfnamefont {G.}~\bibnamefont {Hallinan}}, \bibinfo {author} {\bibfnamefont {V.}~\bibnamefont {Ravi}}, \bibinfo {author} {\bibfnamefont {S.}~\bibnamefont {Weinreb}}, \bibinfo {author} {\bibfnamefont {J.}~\bibnamefont {Kocz}}, \bibinfo {author} {\bibfnamefont {Y.}~\bibnamefont {Huang}}, \bibinfo {author} {\bibfnamefont {D.}~\bibnamefont {Woody}}, \bibinfo {author} {\bibfnamefont {J.}~\bibnamefont {Lamb}}, \bibinfo {author} {\bibfnamefont {L.}~\bibnamefont {D’Addario}}, \bibinfo {author} {\bibfnamefont {M.}~\bibnamefont {Catha}}, \bibinfo {author} {\bibfnamefont {J.}~\bibnamefont {Shi}}, \emph {et~al.},\ }\href@noop {} {\bibfield  {journal} {\bibinfo  {journal} {arXiv preprint arXiv:1907.07648}\ } (\bibinfo {year} {2019})}\BibitemShut {NoStop}%
\bibitem [{\citenamefont {Lin}\ \emph {et~al.}(2022)\citenamefont {Lin}, \citenamefont {Lin}, \citenamefont {Li}, \citenamefont {Tseng}, \citenamefont {Jiang}, \citenamefont {Wang}, \citenamefont {Cheng}, \citenamefont {Pen}, \citenamefont {Chen}, \citenamefont {Chen} \emph {et~al.}}]{lin2022burstt}%
  \BibitemOpen
  \bibfield  {author} {\bibinfo {author} {\bibfnamefont {H.-H.}\ \bibnamefont {Lin}}, \bibinfo {author} {\bibfnamefont {K.-y.}\ \bibnamefont {Lin}}, \bibinfo {author} {\bibfnamefont {C.-T.}\ \bibnamefont {Li}}, \bibinfo {author} {\bibfnamefont {Y.-H.}\ \bibnamefont {Tseng}}, \bibinfo {author} {\bibfnamefont {H.}~\bibnamefont {Jiang}}, \bibinfo {author} {\bibfnamefont {J.-H.}\ \bibnamefont {Wang}}, \bibinfo {author} {\bibfnamefont {J.-C.}\ \bibnamefont {Cheng}}, \bibinfo {author} {\bibfnamefont {U.-L.}\ \bibnamefont {Pen}}, \bibinfo {author} {\bibfnamefont {M.-T.}\ \bibnamefont {Chen}}, \bibinfo {author} {\bibfnamefont {P.}~\bibnamefont {Chen}}, \emph {et~al.},\ }\href@noop {} {\bibfield  {journal} {\bibinfo  {journal} {Publications of the Astronomical Society of the Pacific}\ }\textbf {\bibinfo {volume} {134}},\ \bibinfo {pages} {094106} (\bibinfo {year} {2022})}\BibitemShut {NoStop}%
\bibitem [{\citenamefont {Zhang}(2006)}]{zhang2006testing}%
  \BibitemOpen
  \bibfield  {author} {\bibinfo {author} {\bibfnamefont {P.}~\bibnamefont {Zhang}},\ }\href@noop {} {\bibfield  {journal} {\bibinfo  {journal} {Physical Review D—Particles, Fields, Gravitation, and Cosmology}\ }\textbf {\bibinfo {volume} {73}},\ \bibinfo {pages} {123504} (\bibinfo {year} {2006})}\BibitemShut {NoStop}%
\bibitem [{\citenamefont {Abbott}\ \emph {et~al.}(2017)\citenamefont {Abbott}, \citenamefont {Abbott}, \citenamefont {Abbott}, \citenamefont {Acernese}, \citenamefont {Ackley}, \citenamefont {Adams}, \citenamefont {Adams}, \citenamefont {Addesso}, \citenamefont {Adhikari}, \citenamefont {Adya} \emph {et~al.}}]{abbott2017gw170817}%
  \BibitemOpen
  \bibfield  {author} {\bibinfo {author} {\bibfnamefont {B.~P.}\ \bibnamefont {Abbott}}, \bibinfo {author} {\bibfnamefont {R.}~\bibnamefont {Abbott}}, \bibinfo {author} {\bibfnamefont {T.~D.}\ \bibnamefont {Abbott}}, \bibinfo {author} {\bibfnamefont {F.}~\bibnamefont {Acernese}}, \bibinfo {author} {\bibfnamefont {K.}~\bibnamefont {Ackley}}, \bibinfo {author} {\bibfnamefont {C.}~\bibnamefont {Adams}}, \bibinfo {author} {\bibfnamefont {T.}~\bibnamefont {Adams}}, \bibinfo {author} {\bibfnamefont {P.}~\bibnamefont {Addesso}}, \bibinfo {author} {\bibfnamefont {R.~X.}\ \bibnamefont {Adhikari}}, \bibinfo {author} {\bibfnamefont {V.~B.}\ \bibnamefont {Adya}}, \emph {et~al.},\ }\href@noop {} {\bibfield  {journal} {\bibinfo  {journal} {Physical review letters}\ }\textbf {\bibinfo {volume} {119}},\ \bibinfo {pages} {161101} (\bibinfo {year} {2017})}\BibitemShut {NoStop}%
\bibitem [{\citenamefont {Ezquiaga}\ and\ \citenamefont {Zumalac{\'a}rregui}(2017)}]{ezquiaga2017dark}%
  \BibitemOpen
  \bibfield  {author} {\bibinfo {author} {\bibfnamefont {J.~M.}\ \bibnamefont {Ezquiaga}}\ and\ \bibinfo {author} {\bibfnamefont {M.}~\bibnamefont {Zumalac{\'a}rregui}},\ }\href@noop {} {\bibfield  {journal} {\bibinfo  {journal} {Physical review letters}\ }\textbf {\bibinfo {volume} {119}},\ \bibinfo {pages} {251304} (\bibinfo {year} {2017})}\BibitemShut {NoStop}%
\bibitem [{\citenamefont {Creminelli}\ and\ \citenamefont {Vernizzi}(2017)}]{creminelli2017dark}%
  \BibitemOpen
  \bibfield  {author} {\bibinfo {author} {\bibfnamefont {P.}~\bibnamefont {Creminelli}}\ and\ \bibinfo {author} {\bibfnamefont {F.}~\bibnamefont {Vernizzi}},\ }\href@noop {} {\bibfield  {journal} {\bibinfo  {journal} {Physical review letters}\ }\textbf {\bibinfo {volume} {119}},\ \bibinfo {pages} {251302} (\bibinfo {year} {2017})}\BibitemShut {NoStop}%
\bibitem [{\citenamefont {Baker}\ \emph {et~al.}(2017)\citenamefont {Baker}, \citenamefont {Bellini}, \citenamefont {Ferreira}, \citenamefont {Lagos}, \citenamefont {Noller},\ and\ \citenamefont {Sawicki}}]{baker2017strong}%
  \BibitemOpen
  \bibfield  {author} {\bibinfo {author} {\bibfnamefont {T.}~\bibnamefont {Baker}}, \bibinfo {author} {\bibfnamefont {E.}~\bibnamefont {Bellini}}, \bibinfo {author} {\bibfnamefont {P.~G.}\ \bibnamefont {Ferreira}}, \bibinfo {author} {\bibfnamefont {M.}~\bibnamefont {Lagos}}, \bibinfo {author} {\bibfnamefont {J.}~\bibnamefont {Noller}},\ and\ \bibinfo {author} {\bibfnamefont {I.}~\bibnamefont {Sawicki}},\ }\href@noop {} {\bibfield  {journal} {\bibinfo  {journal} {Physical review letters}\ }\textbf {\bibinfo {volume} {119}},\ \bibinfo {pages} {251301} (\bibinfo {year} {2017})}\BibitemShut {NoStop}%
\bibitem [{\citenamefont {Jeong}\ \emph {et~al.}(2012)\citenamefont {Jeong}, \citenamefont {Schmidt},\ and\ \citenamefont {Hirata}}]{jeong2012large}%
  \BibitemOpen
  \bibfield  {author} {\bibinfo {author} {\bibfnamefont {D.}~\bibnamefont {Jeong}}, \bibinfo {author} {\bibfnamefont {F.}~\bibnamefont {Schmidt}},\ and\ \bibinfo {author} {\bibfnamefont {C.~M.}\ \bibnamefont {Hirata}},\ }\href@noop {} {\bibfield  {journal} {\bibinfo  {journal} {Physical Review D—Particles, Fields, Gravitation, and Cosmology}\ }\textbf {\bibinfo {volume} {85}},\ \bibinfo {pages} {023504} (\bibinfo {year} {2012})}\BibitemShut {NoStop}%
\bibitem [{\citenamefont {Kilbinger}(2015)}]{kilbinger2015cosmology}%
  \BibitemOpen
  \bibfield  {author} {\bibinfo {author} {\bibfnamefont {M.}~\bibnamefont {Kilbinger}},\ }\href@noop {} {\bibfield  {journal} {\bibinfo  {journal} {Reports on Progress in Physics}\ }\textbf {\bibinfo {volume} {78}},\ \bibinfo {pages} {086901} (\bibinfo {year} {2015})}\BibitemShut {NoStop}%
\bibitem [{\citenamefont {Sun}\ \emph {et~al.}(2023)\citenamefont {Sun}, \citenamefont {Zhang}, \citenamefont {Dong}, \citenamefont {Yao}, \citenamefont {Shan}, \citenamefont {Jullo}, \citenamefont {Kneib},\ and\ \citenamefont {Yin}}]{sun2023unbiased}%
  \BibitemOpen
  \bibfield  {author} {\bibinfo {author} {\bibfnamefont {Z.}~\bibnamefont {Sun}}, \bibinfo {author} {\bibfnamefont {P.}~\bibnamefont {Zhang}}, \bibinfo {author} {\bibfnamefont {F.}~\bibnamefont {Dong}}, \bibinfo {author} {\bibfnamefont {J.}~\bibnamefont {Yao}}, \bibinfo {author} {\bibfnamefont {H.}~\bibnamefont {Shan}}, \bibinfo {author} {\bibfnamefont {E.}~\bibnamefont {Jullo}}, \bibinfo {author} {\bibfnamefont {J.-P.}\ \bibnamefont {Kneib}},\ and\ \bibinfo {author} {\bibfnamefont {B.}~\bibnamefont {Yin}},\ }\href@noop {} {\bibfield  {journal} {\bibinfo  {journal} {The Astrophysical Journal Supplement Series}\ }\textbf {\bibinfo {volume} {267}},\ \bibinfo {pages} {21} (\bibinfo {year} {2023})}\BibitemShut {NoStop}%
\bibitem [{\citenamefont {Bonoli}\ and\ \citenamefont {Pen}(2009)}]{bonoli2009halo}%
  \BibitemOpen
  \bibfield  {author} {\bibinfo {author} {\bibfnamefont {S.}~\bibnamefont {Bonoli}}\ and\ \bibinfo {author} {\bibfnamefont {U.-L.}\ \bibnamefont {Pen}},\ }\href@noop {} {\bibfield  {journal} {\bibinfo  {journal} {Monthly Notices of the Royal Astronomical Society}\ }\textbf {\bibinfo {volume} {396}},\ \bibinfo {pages} {1610} (\bibinfo {year} {2009})}\BibitemShut {NoStop}%
\bibitem [{\citenamefont {Seljak}\ \emph {et~al.}(2009)\citenamefont {Seljak}, \citenamefont {Hamaus},\ and\ \citenamefont {Desjacques}}]{seljak2009suppress}%
  \BibitemOpen
  \bibfield  {author} {\bibinfo {author} {\bibfnamefont {U.}~\bibnamefont {Seljak}}, \bibinfo {author} {\bibfnamefont {N.}~\bibnamefont {Hamaus}},\ and\ \bibinfo {author} {\bibfnamefont {V.}~\bibnamefont {Desjacques}},\ }\href@noop {} {\bibfield  {journal} {\bibinfo  {journal} {Physical Review Letters}\ }\textbf {\bibinfo {volume} {103}},\ \bibinfo {pages} {091303} (\bibinfo {year} {2009})}\BibitemShut {NoStop}%
\bibitem [{\citenamefont {Hamaus}\ \emph {et~al.}(2010)\citenamefont {Hamaus}, \citenamefont {Seljak}, \citenamefont {Desjacques}, \citenamefont {Smith},\ and\ \citenamefont {Baldauf}}]{hamaus2010minimizing}%
  \BibitemOpen
  \bibfield  {author} {\bibinfo {author} {\bibfnamefont {N.}~\bibnamefont {Hamaus}}, \bibinfo {author} {\bibfnamefont {U.}~\bibnamefont {Seljak}}, \bibinfo {author} {\bibfnamefont {V.}~\bibnamefont {Desjacques}}, \bibinfo {author} {\bibfnamefont {R.~E.}\ \bibnamefont {Smith}},\ and\ \bibinfo {author} {\bibfnamefont {T.}~\bibnamefont {Baldauf}},\ }\href@noop {} {\bibfield  {journal} {\bibinfo  {journal} {Physical Review D}\ }\textbf {\bibinfo {volume} {82}},\ \bibinfo {pages} {043515} (\bibinfo {year} {2010})}\BibitemShut {NoStop}%
\bibitem [{\citenamefont {Cai}\ \emph {et~al.}(2011)\citenamefont {Cai}, \citenamefont {Bernstein},\ and\ \citenamefont {Sheth}}]{cai2011optimal}%
  \BibitemOpen
  \bibfield  {author} {\bibinfo {author} {\bibfnamefont {Y.-C.}\ \bibnamefont {Cai}}, \bibinfo {author} {\bibfnamefont {G.}~\bibnamefont {Bernstein}},\ and\ \bibinfo {author} {\bibfnamefont {R.~K.}\ \bibnamefont {Sheth}},\ }\href@noop {} {\bibfield  {journal} {\bibinfo  {journal} {Monthly Notices of the Royal Astronomical Society}\ }\textbf {\bibinfo {volume} {412}},\ \bibinfo {pages} {995} (\bibinfo {year} {2011})}\BibitemShut {NoStop}%
\bibitem [{\citenamefont {Liu}\ \emph {et~al.}(2021)\citenamefont {Liu}, \citenamefont {Yu},\ and\ \citenamefont {Li}}]{liu2021biased}%
  \BibitemOpen
  \bibfield  {author} {\bibinfo {author} {\bibfnamefont {Y.}~\bibnamefont {Liu}}, \bibinfo {author} {\bibfnamefont {Y.}~\bibnamefont {Yu}},\ and\ \bibinfo {author} {\bibfnamefont {B.}~\bibnamefont {Li}},\ }\href@noop {} {\bibfield  {journal} {\bibinfo  {journal} {The Astrophysical Journal Supplement Series}\ }\textbf {\bibinfo {volume} {254}},\ \bibinfo {pages} {4} (\bibinfo {year} {2021})}\BibitemShut {NoStop}%
\bibitem [{\citenamefont {Zhou}\ and\ \citenamefont {Zhang}(2024)}]{zhou2024parametrization}%
  \BibitemOpen
  \bibfield  {author} {\bibinfo {author} {\bibfnamefont {S.}~\bibnamefont {Zhou}}\ and\ \bibinfo {author} {\bibfnamefont {P.}~\bibnamefont {Zhang}},\ }\href@noop {} {\bibfield  {journal} {\bibinfo  {journal} {Physical Review D}\ }\textbf {\bibinfo {volume} {110}},\ \bibinfo {pages} {123528} (\bibinfo {year} {2024})}\BibitemShut {NoStop}%
\bibitem [{Note1()}]{Note1}%
  \BibitemOpen
  \bibinfo {note} {For illustration, we consider the estimator $\protect \hat {E}_G\propto \protect \hat {C}^{g\kappa }/(\beta \protect \hat {C}^{gg})$ in Ref.~\cite {pullen2015probing} designed for projected fields. In the approximation of narrow redshift bin, there are $\protect \hat {C}^{g\kappa }\propto P_{gm}=b^D P_{mm}$ and $\protect \hat {C}^{gg}\propto P_{gg} = (b^S)^2 P_{mm}$. The deterministic bias $b^D$ differs from the stochastic bias $b^S$ in the presence of galaxy stochasticity. Meanwhile, the redshift-space distortion parameter is $\beta = f/b^D$, leading to $\protect \hat {E}_G\propto (b^D/b^S)^2 = r^2_{gm} $. Consequently, the $E_G$ is suppressed by a factor $r^2_{gm}<1$, which is the cross-correlation coefficient between the underlying matter and the galaxy clustering.}\BibitemShut {Stop}%
\bibitem [{Note2()}]{Note2}%
  \BibitemOpen
  \bibinfo {note} {The shot noise contribution $\sigma ^2_{\protect \rm DM} = \sigma ^2_{\protect \rm host} + \sigma ^2_{\protect \mathcal {D}}+ \sigma ^2_{\protect \rm MW}$ in the power spectrum of DM is not properly accounted in some forecast works, such as Ref.~\cite {neumann2024fast, shirasaki2022probing}, where they only consider the subdominant host-galaxy contribution $\sigma ^2_{\protect \rm host}$ and result in over-optimistic results. The cosmic DM $\sigma _{\protect \mathcal {D}}^2$ in high redshift FRB can be significantly larger than the host-galaxy contribution. Therefore, the shot noise can dominate the measurement $\protect \hat {C}^{{\protect \mathcal {D}}{\protect \mathcal {D}}}_\ell = C^{{\protect \mathcal {D}}{\protect \mathcal {D}}}_\ell + \sigma ^2_{DM}\protect \,/\protect \bar {n}_{\protect \rm FRB}$ on scale $\ell \gtrsim 10\sim 100$ for typical FRB number $10^4\sim 10^5$ in usual estimation \cite {madhavacheril2019cosmology, sharma2025probing}.}\BibitemShut {Stop}%
\bibitem [{\citenamefont {Ivezi{\'c}}\ \emph {et~al.}(2019)\citenamefont {Ivezi{\'c}}, \citenamefont {Kahn}, \citenamefont {Tyson}, \citenamefont {Abel}, \citenamefont {Acosta}, \citenamefont {Allsman}, \citenamefont {Alonso}, \citenamefont {AlSayyad}, \citenamefont {Anderson}, \citenamefont {Andrew} \emph {et~al.}}]{ivezic2019lsst}%
  \BibitemOpen
  \bibfield  {author} {\bibinfo {author} {\bibfnamefont {{\v{Z}}.}~\bibnamefont {Ivezi{\'c}}}, \bibinfo {author} {\bibfnamefont {S.~M.}\ \bibnamefont {Kahn}}, \bibinfo {author} {\bibfnamefont {J.~A.}\ \bibnamefont {Tyson}}, \bibinfo {author} {\bibfnamefont {B.}~\bibnamefont {Abel}}, \bibinfo {author} {\bibfnamefont {E.}~\bibnamefont {Acosta}}, \bibinfo {author} {\bibfnamefont {R.}~\bibnamefont {Allsman}}, \bibinfo {author} {\bibfnamefont {D.}~\bibnamefont {Alonso}}, \bibinfo {author} {\bibfnamefont {Y.}~\bibnamefont {AlSayyad}}, \bibinfo {author} {\bibfnamefont {S.~F.}\ \bibnamefont {Anderson}}, \bibinfo {author} {\bibfnamefont {J.}~\bibnamefont {Andrew}}, \emph {et~al.},\ }\href@noop {} {\bibfield  {journal} {\bibinfo  {journal} {The Astrophysical Journal}\ }\textbf {\bibinfo {volume} {873}},\ \bibinfo {pages} {111} (\bibinfo {year} {2019})}\BibitemShut {NoStop}%
\bibitem [{\citenamefont {Adame}\ \emph {et~al.}(2024)\citenamefont {Adame}, \citenamefont {Aguilar}, \citenamefont {Ahlen}, \citenamefont {Alam}, \citenamefont {Aldering}, \citenamefont {Alexander}, \citenamefont {Alfarsy}, \citenamefont {Prieto}, \citenamefont {Alvarez}, \citenamefont {Alves} \emph {et~al.}}]{adame2024validation}%
  \BibitemOpen
  \bibfield  {author} {\bibinfo {author} {\bibfnamefont {A.}~\bibnamefont {Adame}}, \bibinfo {author} {\bibfnamefont {J.}~\bibnamefont {Aguilar}}, \bibinfo {author} {\bibfnamefont {S.}~\bibnamefont {Ahlen}}, \bibinfo {author} {\bibfnamefont {S.}~\bibnamefont {Alam}}, \bibinfo {author} {\bibfnamefont {G.}~\bibnamefont {Aldering}}, \bibinfo {author} {\bibfnamefont {D.}~\bibnamefont {Alexander}}, \bibinfo {author} {\bibfnamefont {R.}~\bibnamefont {Alfarsy}}, \bibinfo {author} {\bibfnamefont {C.~A.}\ \bibnamefont {Prieto}}, \bibinfo {author} {\bibfnamefont {M.}~\bibnamefont {Alvarez}}, \bibinfo {author} {\bibfnamefont {O.}~\bibnamefont {Alves}}, \emph {et~al.},\ }\href@noop {} {\bibfield  {journal} {\bibinfo  {journal} {The Astronomical Journal}\ }\textbf {\bibinfo {volume} {167}},\ \bibinfo {pages} {62} (\bibinfo {year} {2024})}\BibitemShut {NoStop}%
\bibitem [{\citenamefont {Chang}\ \emph {et~al.}(2013)\citenamefont {Chang}, \citenamefont {Jarvis}, \citenamefont {Jain}, \citenamefont {Kahn}, \citenamefont {Kirkby}, \citenamefont {Connolly}, \citenamefont {Krughoff}, \citenamefont {Peng},\ and\ \citenamefont {Peterson}}]{chang2013effective}%
  \BibitemOpen
  \bibfield  {author} {\bibinfo {author} {\bibfnamefont {C.}~\bibnamefont {Chang}}, \bibinfo {author} {\bibfnamefont {M.}~\bibnamefont {Jarvis}}, \bibinfo {author} {\bibfnamefont {B.}~\bibnamefont {Jain}}, \bibinfo {author} {\bibfnamefont {S.}~\bibnamefont {Kahn}}, \bibinfo {author} {\bibfnamefont {D.}~\bibnamefont {Kirkby}}, \bibinfo {author} {\bibfnamefont {A.}~\bibnamefont {Connolly}}, \bibinfo {author} {\bibfnamefont {S.}~\bibnamefont {Krughoff}}, \bibinfo {author} {\bibfnamefont {E.-H.}\ \bibnamefont {Peng}},\ and\ \bibinfo {author} {\bibfnamefont {J.}~\bibnamefont {Peterson}},\ }\href@noop {} {\bibfield  {journal} {\bibinfo  {journal} {Monthly Notices of the Royal Astronomical Society}\ }\textbf {\bibinfo {volume} {434}},\ \bibinfo {pages} {2121} (\bibinfo {year} {2013})}\BibitemShut {NoStop}%
\bibitem [{\citenamefont {Rafiei-Ravandi}\ \emph {et~al.}(2021)\citenamefont {Rafiei-Ravandi}, \citenamefont {Smith}, \citenamefont {Li}, \citenamefont {Masui}, \citenamefont {Josephy}, \citenamefont {Dobbs}, \citenamefont {Lang}, \citenamefont {Bhardwaj}, \citenamefont {Patel}, \citenamefont {Bandura} \emph {et~al.}}]{rafiei2021chime}%
  \BibitemOpen
  \bibfield  {author} {\bibinfo {author} {\bibfnamefont {M.}~\bibnamefont {Rafiei-Ravandi}}, \bibinfo {author} {\bibfnamefont {K.~M.}\ \bibnamefont {Smith}}, \bibinfo {author} {\bibfnamefont {D.}~\bibnamefont {Li}}, \bibinfo {author} {\bibfnamefont {K.~W.}\ \bibnamefont {Masui}}, \bibinfo {author} {\bibfnamefont {A.}~\bibnamefont {Josephy}}, \bibinfo {author} {\bibfnamefont {M.}~\bibnamefont {Dobbs}}, \bibinfo {author} {\bibfnamefont {D.}~\bibnamefont {Lang}}, \bibinfo {author} {\bibfnamefont {M.}~\bibnamefont {Bhardwaj}}, \bibinfo {author} {\bibfnamefont {C.}~\bibnamefont {Patel}}, \bibinfo {author} {\bibfnamefont {K.}~\bibnamefont {Bandura}}, \emph {et~al.},\ }\href@noop {} {\bibfield  {journal} {\bibinfo  {journal} {The Astrophysical Journal}\ }\textbf {\bibinfo {volume} {922}},\ \bibinfo {pages} {42} (\bibinfo {year} {2021})}\BibitemShut {NoStop}%
\bibitem [{\citenamefont {Neumann}\ \emph {et~al.}(2024)\citenamefont {Neumann}, \citenamefont {Reischke}, \citenamefont {Hagstotz},\ and\ \citenamefont {Hildebrandt}}]{neumann2024fast}%
  \BibitemOpen
  \bibfield  {author} {\bibinfo {author} {\bibfnamefont {D.}~\bibnamefont {Neumann}}, \bibinfo {author} {\bibfnamefont {R.}~\bibnamefont {Reischke}}, \bibinfo {author} {\bibfnamefont {S.}~\bibnamefont {Hagstotz}},\ and\ \bibinfo {author} {\bibfnamefont {H.}~\bibnamefont {Hildebrandt}},\ }\href@noop {} {\bibfield  {journal} {\bibinfo  {journal} {arXiv preprint arXiv:2409.11163}\ } (\bibinfo {year} {2024})}\BibitemShut {NoStop}%
\bibitem [{\citenamefont {Schaller}\ \emph {et~al.}(2025)\citenamefont {Schaller}, \citenamefont {Schaye}, \citenamefont {Kugel}, \citenamefont {Broxterman},\ and\ \citenamefont {van Daalen}}]{schaller2025flamingo}%
  \BibitemOpen
  \bibfield  {author} {\bibinfo {author} {\bibfnamefont {M.}~\bibnamefont {Schaller}}, \bibinfo {author} {\bibfnamefont {J.}~\bibnamefont {Schaye}}, \bibinfo {author} {\bibfnamefont {R.}~\bibnamefont {Kugel}}, \bibinfo {author} {\bibfnamefont {J.~C.}\ \bibnamefont {Broxterman}},\ and\ \bibinfo {author} {\bibfnamefont {M.~P.}\ \bibnamefont {van Daalen}},\ }\href@noop {} {\bibfield  {journal} {\bibinfo  {journal} {Monthly Notices of the Royal Astronomical Society}\ }\textbf {\bibinfo {volume} {539}},\ \bibinfo {pages} {1337} (\bibinfo {year} {2025})}\BibitemShut {NoStop}%
\bibitem [{\citenamefont {Siegel}\ \emph {et~al.}(2025)\citenamefont {Siegel}, \citenamefont {Bigwood}, \citenamefont {Amon}, \citenamefont {McCullough}, \citenamefont {Yamamoto}, \citenamefont {McCarthy}, \citenamefont {Schaller}, \citenamefont {Schneider},\ and\ \citenamefont {Schaye}}]{siegel2025suppression}%
  \BibitemOpen
  \bibfield  {author} {\bibinfo {author} {\bibfnamefont {J.}~\bibnamefont {Siegel}}, \bibinfo {author} {\bibfnamefont {L.}~\bibnamefont {Bigwood}}, \bibinfo {author} {\bibfnamefont {A.}~\bibnamefont {Amon}}, \bibinfo {author} {\bibfnamefont {J.}~\bibnamefont {McCullough}}, \bibinfo {author} {\bibfnamefont {M.}~\bibnamefont {Yamamoto}}, \bibinfo {author} {\bibfnamefont {I.~G.}\ \bibnamefont {McCarthy}}, \bibinfo {author} {\bibfnamefont {M.}~\bibnamefont {Schaller}}, \bibinfo {author} {\bibfnamefont {A.}~\bibnamefont {Schneider}},\ and\ \bibinfo {author} {\bibfnamefont {J.}~\bibnamefont {Schaye}},\ }\href@noop {} {\bibfield  {journal} {\bibinfo  {journal} {arXiv preprint arXiv:2512.02954}\ } (\bibinfo {year} {2025})}\BibitemShut {NoStop}%
\bibitem [{Note3()}]{Note3}%
  \BibitemOpen
  \bibinfo {note} {We neglect the sub-dominating Milky Way contribution, and integrate the cosmic contribution $\sigma _{\protect \mathcal {D}}$ up to $\ell =3000$.}\BibitemShut {Stop}%
\bibitem [{\citenamefont {Aghanim}\ \emph {et~al.}(2020)\citenamefont {Aghanim} \emph {et~al.}}]{aghanim2020planck}%
  \BibitemOpen
  \bibfield  {author} {\bibinfo {author} {\bibfnamefont {N.}~\bibnamefont {Aghanim}} \emph {et~al.},\ }\href@noop {} {\bibfield  {journal} {\bibinfo  {journal} {Astron. Astrophys}\ }\textbf {\bibinfo {volume} {641}},\ \bibinfo {pages} {A6} (\bibinfo {year} {2020})}\BibitemShut {NoStop}%
\bibitem [{\citenamefont {Louis}\ \emph {et~al.}(2025)\citenamefont {Louis}, \citenamefont {La~Posta}, \citenamefont {Atkins}, \citenamefont {Jense}, \citenamefont {Abril-Cabezas}, \citenamefont {Addison}, \citenamefont {Ade}, \citenamefont {Aiola}, \citenamefont {Alford}, \citenamefont {Alonso} \emph {et~al.}}]{louis2025atacama}%
  \BibitemOpen
  \bibfield  {author} {\bibinfo {author} {\bibfnamefont {T.}~\bibnamefont {Louis}}, \bibinfo {author} {\bibfnamefont {A.}~\bibnamefont {La~Posta}}, \bibinfo {author} {\bibfnamefont {Z.}~\bibnamefont {Atkins}}, \bibinfo {author} {\bibfnamefont {H.~T.}\ \bibnamefont {Jense}}, \bibinfo {author} {\bibfnamefont {I.}~\bibnamefont {Abril-Cabezas}}, \bibinfo {author} {\bibfnamefont {G.~E.}\ \bibnamefont {Addison}}, \bibinfo {author} {\bibfnamefont {P.~A.}\ \bibnamefont {Ade}}, \bibinfo {author} {\bibfnamefont {S.}~\bibnamefont {Aiola}}, \bibinfo {author} {\bibfnamefont {T.}~\bibnamefont {Alford}}, \bibinfo {author} {\bibfnamefont {D.}~\bibnamefont {Alonso}}, \emph {et~al.},\ }\href@noop {} {\bibfield  {journal} {\bibinfo  {journal} {arXiv preprint arXiv:2503.14452}\ } (\bibinfo {year} {2025})}\BibitemShut {NoStop}%
\bibitem [{\citenamefont {Vallenari}\ \emph {et~al.}(2023)\citenamefont {Vallenari}, \citenamefont {Brown}, \citenamefont {Prusti}, \citenamefont {De~Bruijne}, \citenamefont {Arenou}, \citenamefont {Babusiaux}, \citenamefont {Biermann}, \citenamefont {Creevey}, \citenamefont {Ducourant}, \citenamefont {Evans} \emph {et~al.}}]{vallenari2023gaia}%
  \BibitemOpen
  \bibfield  {author} {\bibinfo {author} {\bibfnamefont {A.}~\bibnamefont {Vallenari}}, \bibinfo {author} {\bibfnamefont {A.~G.}\ \bibnamefont {Brown}}, \bibinfo {author} {\bibfnamefont {T.}~\bibnamefont {Prusti}}, \bibinfo {author} {\bibfnamefont {J.~H.}\ \bibnamefont {De~Bruijne}}, \bibinfo {author} {\bibfnamefont {F.}~\bibnamefont {Arenou}}, \bibinfo {author} {\bibfnamefont {C.}~\bibnamefont {Babusiaux}}, \bibinfo {author} {\bibfnamefont {M.}~\bibnamefont {Biermann}}, \bibinfo {author} {\bibfnamefont {O.~L.}\ \bibnamefont {Creevey}}, \bibinfo {author} {\bibfnamefont {C.}~\bibnamefont {Ducourant}}, \bibinfo {author} {\bibfnamefont {D.~W.}\ \bibnamefont {Evans}}, \emph {et~al.},\ }\href@noop {} {\bibfield  {journal} {\bibinfo  {journal} {Astronomy \& Astrophysics}\ }\textbf {\bibinfo {volume} {674}},\ \bibinfo {pages} {A1} (\bibinfo {year} {2023})}\BibitemShut {NoStop}%
\bibitem [{\citenamefont {Lutsenko}\ \emph {et~al.}(2025)\citenamefont {Lutsenko}, \citenamefont {Carraro}, \citenamefont {Korchagin}, \citenamefont {Tkachenko},\ and\ \citenamefont {Vieira}}]{lutsenko2025counting}%
  \BibitemOpen
  \bibfield  {author} {\bibinfo {author} {\bibfnamefont {A.}~\bibnamefont {Lutsenko}}, \bibinfo {author} {\bibfnamefont {G.}~\bibnamefont {Carraro}}, \bibinfo {author} {\bibfnamefont {V.}~\bibnamefont {Korchagin}}, \bibinfo {author} {\bibfnamefont {R.}~\bibnamefont {Tkachenko}},\ and\ \bibinfo {author} {\bibfnamefont {K.}~\bibnamefont {Vieira}},\ }\href@noop {} {\bibfield  {journal} {\bibinfo  {journal} {The Astrophysical Journal}\ }\textbf {\bibinfo {volume} {990}},\ \bibinfo {pages} {88} (\bibinfo {year} {2025})}\BibitemShut {NoStop}%
\bibitem [{\citenamefont {Pouteau}\ \emph {et~al.}(2022)\citenamefont {Pouteau}, \citenamefont {Motte}, \citenamefont {Nony}, \citenamefont {Galv{\'a}n-Madrid}, \citenamefont {Men’shchikov}, \citenamefont {Bontemps}, \citenamefont {Robitaille}, \citenamefont {Louvet}, \citenamefont {Ginsburg}, \citenamefont {Herpin} \emph {et~al.}}]{pouteau2022alma}%
  \BibitemOpen
  \bibfield  {author} {\bibinfo {author} {\bibfnamefont {Y.}~\bibnamefont {Pouteau}}, \bibinfo {author} {\bibfnamefont {F.}~\bibnamefont {Motte}}, \bibinfo {author} {\bibfnamefont {T.}~\bibnamefont {Nony}}, \bibinfo {author} {\bibfnamefont {R.}~\bibnamefont {Galv{\'a}n-Madrid}}, \bibinfo {author} {\bibfnamefont {A.}~\bibnamefont {Men’shchikov}}, \bibinfo {author} {\bibfnamefont {S.}~\bibnamefont {Bontemps}}, \bibinfo {author} {\bibfnamefont {J.-F.}\ \bibnamefont {Robitaille}}, \bibinfo {author} {\bibfnamefont {F.}~\bibnamefont {Louvet}}, \bibinfo {author} {\bibfnamefont {A.}~\bibnamefont {Ginsburg}}, \bibinfo {author} {\bibfnamefont {F.}~\bibnamefont {Herpin}}, \emph {et~al.},\ }\href@noop {} {\bibfield  {journal} {\bibinfo  {journal} {Astronomy \& Astrophysics}\ }\textbf {\bibinfo {volume} {664}},\ \bibinfo {pages} {A26} (\bibinfo {year} {2022})}\BibitemShut {NoStop}%
\bibitem [{\citenamefont {Pouteau}\ \emph {et~al.}(2023)\citenamefont {Pouteau}, \citenamefont {Motte}, \citenamefont {Nony}, \citenamefont {Gonz{\'a}lez}, \citenamefont {Joncour}, \citenamefont {Robitaille}, \citenamefont {Busquet}, \citenamefont {Galv{\'a}n-Madrid}, \citenamefont {Gusdorf}, \citenamefont {Hennebelle} \emph {et~al.}}]{pouteau2023alma}%
  \BibitemOpen
  \bibfield  {author} {\bibinfo {author} {\bibfnamefont {Y.}~\bibnamefont {Pouteau}}, \bibinfo {author} {\bibfnamefont {F.}~\bibnamefont {Motte}}, \bibinfo {author} {\bibfnamefont {T.}~\bibnamefont {Nony}}, \bibinfo {author} {\bibfnamefont {M.}~\bibnamefont {Gonz{\'a}lez}}, \bibinfo {author} {\bibfnamefont {I.}~\bibnamefont {Joncour}}, \bibinfo {author} {\bibfnamefont {J.-F.}\ \bibnamefont {Robitaille}}, \bibinfo {author} {\bibfnamefont {G.}~\bibnamefont {Busquet}}, \bibinfo {author} {\bibfnamefont {R.}~\bibnamefont {Galv{\'a}n-Madrid}}, \bibinfo {author} {\bibfnamefont {A.}~\bibnamefont {Gusdorf}}, \bibinfo {author} {\bibfnamefont {P.}~\bibnamefont {Hennebelle}}, \emph {et~al.},\ }\href@noop {} {\bibfield  {journal} {\bibinfo  {journal} {Astronomy \& Astrophysics}\ }\textbf {\bibinfo {volume} {674}},\ \bibinfo {pages} {A76} (\bibinfo {year} {2023})}\BibitemShut {NoStop}%
\bibitem [{\citenamefont {Xu}\ \emph {et~al.}(2025)\citenamefont {Xu}, \citenamefont {Jing}, \citenamefont {Cole}, \citenamefont {Frenk}, \citenamefont {Bose}, \citenamefont {Elbers}, \citenamefont {Wang}, \citenamefont {Wang}, \citenamefont {Moore}, \citenamefont {Aguilar} \emph {et~al.}}]{xu2025pac}%
  \BibitemOpen
  \bibfield  {author} {\bibinfo {author} {\bibfnamefont {K.}~\bibnamefont {Xu}}, \bibinfo {author} {\bibfnamefont {Y.}~\bibnamefont {Jing}}, \bibinfo {author} {\bibfnamefont {S.}~\bibnamefont {Cole}}, \bibinfo {author} {\bibfnamefont {C.}~\bibnamefont {Frenk}}, \bibinfo {author} {\bibfnamefont {S.}~\bibnamefont {Bose}}, \bibinfo {author} {\bibfnamefont {W.}~\bibnamefont {Elbers}}, \bibinfo {author} {\bibfnamefont {W.}~\bibnamefont {Wang}}, \bibinfo {author} {\bibfnamefont {Y.}~\bibnamefont {Wang}}, \bibinfo {author} {\bibfnamefont {S.}~\bibnamefont {Moore}}, \bibinfo {author} {\bibfnamefont {J.}~\bibnamefont {Aguilar}}, \emph {et~al.},\ }\href@noop {} {\bibfield  {journal} {\bibinfo  {journal} {Monthly Notices of the Royal Astronomical Society}\ }\textbf {\bibinfo {volume} {540}},\ \bibinfo {pages} {1635} (\bibinfo {year} {2025})}\BibitemShut {NoStop}%
\bibitem [{\citenamefont {Wang}\ \emph {et~al.}(2025)\citenamefont {Wang}, \citenamefont {Yang}, \citenamefont {Jing}, \citenamefont {Ross}, \citenamefont {Siudek}, \citenamefont {Moustakas}, \citenamefont {Moore}, \citenamefont {Cole}, \citenamefont {Frenk}, \citenamefont {Yu} \emph {et~al.}}]{wang2025luminosity}%
  \BibitemOpen
  \bibfield  {author} {\bibinfo {author} {\bibfnamefont {W.}~\bibnamefont {Wang}}, \bibinfo {author} {\bibfnamefont {X.}~\bibnamefont {Yang}}, \bibinfo {author} {\bibfnamefont {Y.}~\bibnamefont {Jing}}, \bibinfo {author} {\bibfnamefont {A.~J.}\ \bibnamefont {Ross}}, \bibinfo {author} {\bibfnamefont {M.}~\bibnamefont {Siudek}}, \bibinfo {author} {\bibfnamefont {J.}~\bibnamefont {Moustakas}}, \bibinfo {author} {\bibfnamefont {S.~G.}\ \bibnamefont {Moore}}, \bibinfo {author} {\bibfnamefont {S.}~\bibnamefont {Cole}}, \bibinfo {author} {\bibfnamefont {C.}~\bibnamefont {Frenk}}, \bibinfo {author} {\bibfnamefont {J.}~\bibnamefont {Yu}}, \emph {et~al.},\ }\href@noop {} {\bibfield  {journal} {\bibinfo  {journal} {The Astrophysical Journal}\ }\textbf {\bibinfo {volume} {986}},\ \bibinfo {pages} {218} (\bibinfo {year} {2025})}\BibitemShut {NoStop}%
\bibitem [{\citenamefont {Amiri}\ \emph {et~al.}(2022)\citenamefont {Amiri}, \citenamefont {Bandura}, \citenamefont {Boskovic}, \citenamefont {Chen}, \citenamefont {Cliche}, \citenamefont {Deng}, \citenamefont {Denman}, \citenamefont {Dobbs}, \citenamefont {Fandino}, \citenamefont {Foreman} \emph {et~al.}}]{amiri2022overview}%
  \BibitemOpen
  \bibfield  {author} {\bibinfo {author} {\bibfnamefont {M.}~\bibnamefont {Amiri}}, \bibinfo {author} {\bibfnamefont {K.}~\bibnamefont {Bandura}}, \bibinfo {author} {\bibfnamefont {A.}~\bibnamefont {Boskovic}}, \bibinfo {author} {\bibfnamefont {T.}~\bibnamefont {Chen}}, \bibinfo {author} {\bibfnamefont {J.-F.}\ \bibnamefont {Cliche}}, \bibinfo {author} {\bibfnamefont {M.}~\bibnamefont {Deng}}, \bibinfo {author} {\bibfnamefont {N.}~\bibnamefont {Denman}}, \bibinfo {author} {\bibfnamefont {M.}~\bibnamefont {Dobbs}}, \bibinfo {author} {\bibfnamefont {M.}~\bibnamefont {Fandino}}, \bibinfo {author} {\bibfnamefont {S.}~\bibnamefont {Foreman}}, \emph {et~al.},\ }\href@noop {} {\bibfield  {journal} {\bibinfo  {journal} {The Astrophysical Journal Supplement Series}\ }\textbf {\bibinfo {volume} {261}},\ \bibinfo {pages} {29} (\bibinfo {year} {2022})}\BibitemShut {NoStop}%
\bibitem [{\citenamefont {Braun}\ \emph {et~al.}(2015)\citenamefont {Braun}, \citenamefont {Bourke}, \citenamefont {Green}, \citenamefont {Keane},\ and\ \citenamefont {Wagg}}]{braun2015advancing}%
  \BibitemOpen
  \bibfield  {author} {\bibinfo {author} {\bibfnamefont {R.}~\bibnamefont {Braun}}, \bibinfo {author} {\bibfnamefont {T.~L.}\ \bibnamefont {Bourke}}, \bibinfo {author} {\bibfnamefont {J.~A.}\ \bibnamefont {Green}}, \bibinfo {author} {\bibfnamefont {E.}~\bibnamefont {Keane}},\ and\ \bibinfo {author} {\bibfnamefont {J.}~\bibnamefont {Wagg}},\ }in\ \href@noop {} {\emph {\bibinfo {booktitle} {Advancing Astrophysics with the Square Kilometre Array}}},\ Vol.\ \bibinfo {volume} {215}\ (\bibinfo {organization} {Sissa Medialab},\ \bibinfo {year} {2015})\ p.\ \bibinfo {pages} {174}\BibitemShut {NoStop}%
\bibitem [{\citenamefont {Masui}\ \emph {et~al.}(2013)\citenamefont {Masui}, \citenamefont {Switzer}, \citenamefont {Banavar}, \citenamefont {Bandura}, \citenamefont {Blake}, \citenamefont {Calin}, \citenamefont {Chang}, \citenamefont {Chen}, \citenamefont {Li}, \citenamefont {Liao} \emph {et~al.}}]{masui2013measurement}%
  \BibitemOpen
  \bibfield  {author} {\bibinfo {author} {\bibfnamefont {K.}~\bibnamefont {Masui}}, \bibinfo {author} {\bibfnamefont {E.}~\bibnamefont {Switzer}}, \bibinfo {author} {\bibfnamefont {N.}~\bibnamefont {Banavar}}, \bibinfo {author} {\bibfnamefont {K.}~\bibnamefont {Bandura}}, \bibinfo {author} {\bibfnamefont {C.}~\bibnamefont {Blake}}, \bibinfo {author} {\bibfnamefont {L.-M.}\ \bibnamefont {Calin}}, \bibinfo {author} {\bibfnamefont {T.-C.}\ \bibnamefont {Chang}}, \bibinfo {author} {\bibfnamefont {X.}~\bibnamefont {Chen}}, \bibinfo {author} {\bibfnamefont {Y.-C.}\ \bibnamefont {Li}}, \bibinfo {author} {\bibfnamefont {Y.-W.}\ \bibnamefont {Liao}}, \emph {et~al.},\ }\href@noop {} {\bibfield  {journal} {\bibinfo  {journal} {The Astrophysical Journal Letters}\ }\textbf {\bibinfo {volume} {763}},\ \bibinfo {pages} {L20} (\bibinfo {year} {2013})}\BibitemShut {NoStop}%
\bibitem [{\citenamefont {Cunnington}\ \emph {et~al.}(2023)\citenamefont {Cunnington}, \citenamefont {Li}, \citenamefont {Santos}, \citenamefont {Wang}, \citenamefont {Carucci}, \citenamefont {Irfan}, \citenamefont {Pourtsidou}, \citenamefont {Spinelli}, \citenamefont {Wolz}, \citenamefont {Soares} \emph {et~al.}}]{cunnington2023h}%
  \BibitemOpen
  \bibfield  {author} {\bibinfo {author} {\bibfnamefont {S.}~\bibnamefont {Cunnington}}, \bibinfo {author} {\bibfnamefont {Y.}~\bibnamefont {Li}}, \bibinfo {author} {\bibfnamefont {M.~G.}\ \bibnamefont {Santos}}, \bibinfo {author} {\bibfnamefont {J.}~\bibnamefont {Wang}}, \bibinfo {author} {\bibfnamefont {I.~P.}\ \bibnamefont {Carucci}}, \bibinfo {author} {\bibfnamefont {M.~O.}\ \bibnamefont {Irfan}}, \bibinfo {author} {\bibfnamefont {A.}~\bibnamefont {Pourtsidou}}, \bibinfo {author} {\bibfnamefont {M.}~\bibnamefont {Spinelli}}, \bibinfo {author} {\bibfnamefont {L.}~\bibnamefont {Wolz}}, \bibinfo {author} {\bibfnamefont {P.~S.}\ \bibnamefont {Soares}}, \emph {et~al.},\ }\href@noop {} {\bibfield  {journal} {\bibinfo  {journal} {Monthly Notices of the Royal Astronomical Society}\ }\textbf {\bibinfo {volume} {518}},\ \bibinfo {pages} {6262} (\bibinfo {year} {2023})}\BibitemShut {NoStop}%
\bibitem [{\citenamefont {Amiri}\ \emph {et~al.}(2023)\citenamefont {Amiri}, \citenamefont {Bandura}, \citenamefont {Chen}, \citenamefont {Deng}, \citenamefont {Dobbs}, \citenamefont {Fandino}, \citenamefont {Foreman}, \citenamefont {Halpern}, \citenamefont {Hill}, \citenamefont {Hinshaw} \emph {et~al.}}]{amiri2023detection}%
  \BibitemOpen
  \bibfield  {author} {\bibinfo {author} {\bibfnamefont {M.}~\bibnamefont {Amiri}}, \bibinfo {author} {\bibfnamefont {K.}~\bibnamefont {Bandura}}, \bibinfo {author} {\bibfnamefont {T.}~\bibnamefont {Chen}}, \bibinfo {author} {\bibfnamefont {M.}~\bibnamefont {Deng}}, \bibinfo {author} {\bibfnamefont {M.}~\bibnamefont {Dobbs}}, \bibinfo {author} {\bibfnamefont {M.}~\bibnamefont {Fandino}}, \bibinfo {author} {\bibfnamefont {S.}~\bibnamefont {Foreman}}, \bibinfo {author} {\bibfnamefont {M.}~\bibnamefont {Halpern}}, \bibinfo {author} {\bibfnamefont {A.~S.}\ \bibnamefont {Hill}}, \bibinfo {author} {\bibfnamefont {G.}~\bibnamefont {Hinshaw}}, \emph {et~al.},\ }\href@noop {} {\bibfield  {journal} {\bibinfo  {journal} {The Astrophysical Journal}\ }\textbf {\bibinfo {volume} {947}},\ \bibinfo {pages} {16} (\bibinfo {year} {2023})}\BibitemShut {NoStop}%
\bibitem [{\citenamefont {Amiri}\ \emph {et~al.}(2024)\citenamefont {Amiri}, \citenamefont {Bandura}, \citenamefont {Chakraborty}, \citenamefont {Dobbs}, \citenamefont {Fandino}, \citenamefont {Foreman}, \citenamefont {Gan}, \citenamefont {Halpern}, \citenamefont {Hill}, \citenamefont {Hinshaw} \emph {et~al.}}]{amiri2024detection}%
  \BibitemOpen
  \bibfield  {author} {\bibinfo {author} {\bibfnamefont {M.}~\bibnamefont {Amiri}}, \bibinfo {author} {\bibfnamefont {K.}~\bibnamefont {Bandura}}, \bibinfo {author} {\bibfnamefont {A.}~\bibnamefont {Chakraborty}}, \bibinfo {author} {\bibfnamefont {M.}~\bibnamefont {Dobbs}}, \bibinfo {author} {\bibfnamefont {M.}~\bibnamefont {Fandino}}, \bibinfo {author} {\bibfnamefont {S.}~\bibnamefont {Foreman}}, \bibinfo {author} {\bibfnamefont {H.}~\bibnamefont {Gan}}, \bibinfo {author} {\bibfnamefont {M.}~\bibnamefont {Halpern}}, \bibinfo {author} {\bibfnamefont {A.~S.}\ \bibnamefont {Hill}}, \bibinfo {author} {\bibfnamefont {G.}~\bibnamefont {Hinshaw}}, \emph {et~al.},\ }\href@noop {} {\bibfield  {journal} {\bibinfo  {journal} {The Astrophysical Journal}\ }\textbf {\bibinfo {volume} {963}},\ \bibinfo {pages} {23} (\bibinfo {year} {2024})}\BibitemShut {NoStop}%
\bibitem [{Note4()}]{Note4}%
  \BibitemOpen
  \bibinfo {note} {Besides those based upon Eq. ~(\ref {equ:be_from_bb}), the kinetic Sunyaev–Zel'dovich effect offers another potential pathway to constrain $b_e$ given its sensitivity to all free electrons in diffuse gas, yet its reconstruction such as four-point estimations depends on the template of LSS tracer velocities \cite {smith2017detecting, kumar2025electrons}.}\BibitemShut {Stop}%
\bibitem [{\citenamefont {Springel}\ \emph {et~al.}(2017)\citenamefont {Springel}, \citenamefont {Pakmor}, \citenamefont {Pillepich}, \citenamefont {Weinberger}, \citenamefont {Nelson}, \citenamefont {Hernquist}, \citenamefont {Vogelsberger}, \citenamefont {Genel}, \citenamefont {Torrey}, \citenamefont {Marinacci},\ and\ \citenamefont {Naiman}}]{Springel_2017}%
  \BibitemOpen
  \bibfield  {author} {\bibinfo {author} {\bibfnamefont {V.}~\bibnamefont {Springel}}, \bibinfo {author} {\bibfnamefont {R.}~\bibnamefont {Pakmor}}, \bibinfo {author} {\bibfnamefont {A.}~\bibnamefont {Pillepich}}, \bibinfo {author} {\bibfnamefont {R.}~\bibnamefont {Weinberger}}, \bibinfo {author} {\bibfnamefont {D.}~\bibnamefont {Nelson}}, \bibinfo {author} {\bibfnamefont {L.}~\bibnamefont {Hernquist}}, \bibinfo {author} {\bibfnamefont {M.}~\bibnamefont {Vogelsberger}}, \bibinfo {author} {\bibfnamefont {S.}~\bibnamefont {Genel}}, \bibinfo {author} {\bibfnamefont {P.}~\bibnamefont {Torrey}}, \bibinfo {author} {\bibfnamefont {F.}~\bibnamefont {Marinacci}},\ and\ \bibinfo {author} {\bibfnamefont {J.}~\bibnamefont {Naiman}},\ }\href {https://doi.org/10.1093/mnras/stx3304} {\bibfield  {journal} {\bibinfo  {journal} {Monthly Notices of the Royal Astronomical Society}\ }\textbf {\bibinfo {volume} {475}},\ \bibinfo {pages} {676} (\bibinfo {year} {2017})}\BibitemShut {NoStop}%
\bibitem [{\citenamefont {Nelson}\ \emph {et~al.}(2017)\citenamefont {Nelson}, \citenamefont {Pillepich}, \citenamefont {Springel}, \citenamefont {Weinberger}, \citenamefont {Hernquist}, \citenamefont {Pakmor}, \citenamefont {Genel}, \citenamefont {Torrey}, \citenamefont {Vogelsberger}, \citenamefont {Kauffmann}, \citenamefont {Marinacci},\ and\ \citenamefont {Naiman}}]{Nelson_2017}%
  \BibitemOpen
  \bibfield  {author} {\bibinfo {author} {\bibfnamefont {D.}~\bibnamefont {Nelson}}, \bibinfo {author} {\bibfnamefont {A.}~\bibnamefont {Pillepich}}, \bibinfo {author} {\bibfnamefont {V.}~\bibnamefont {Springel}}, \bibinfo {author} {\bibfnamefont {R.}~\bibnamefont {Weinberger}}, \bibinfo {author} {\bibfnamefont {L.}~\bibnamefont {Hernquist}}, \bibinfo {author} {\bibfnamefont {R.}~\bibnamefont {Pakmor}}, \bibinfo {author} {\bibfnamefont {S.}~\bibnamefont {Genel}}, \bibinfo {author} {\bibfnamefont {P.}~\bibnamefont {Torrey}}, \bibinfo {author} {\bibfnamefont {M.}~\bibnamefont {Vogelsberger}}, \bibinfo {author} {\bibfnamefont {G.}~\bibnamefont {Kauffmann}}, \bibinfo {author} {\bibfnamefont {F.}~\bibnamefont {Marinacci}},\ and\ \bibinfo {author} {\bibfnamefont {J.}~\bibnamefont {Naiman}},\ }\href {https://doi.org/10.1093/mnras/stx3040} {\bibfield  {journal} {\bibinfo  {journal} {Monthly Notices of the Royal Astronomical Society}\ }\textbf {\bibinfo {volume} {475}},\ \bibinfo {pages} {624} (\bibinfo {year} {2017})}\BibitemShut {NoStop}%
\bibitem [{\citenamefont {Pillepich}\ \emph {et~al.}(2017)\citenamefont {Pillepich}, \citenamefont {Nelson}, \citenamefont {Hernquist}, \citenamefont {Springel}, \citenamefont {Pakmor}, \citenamefont {Torrey}, \citenamefont {Weinberger}, \citenamefont {Genel}, \citenamefont {Naiman}, \citenamefont {Marinacci},\ and\ \citenamefont {Vogelsberger}}]{Pillepich_2017}%
  \BibitemOpen
  \bibfield  {author} {\bibinfo {author} {\bibfnamefont {A.}~\bibnamefont {Pillepich}}, \bibinfo {author} {\bibfnamefont {D.}~\bibnamefont {Nelson}}, \bibinfo {author} {\bibfnamefont {L.}~\bibnamefont {Hernquist}}, \bibinfo {author} {\bibfnamefont {V.}~\bibnamefont {Springel}}, \bibinfo {author} {\bibfnamefont {R.}~\bibnamefont {Pakmor}}, \bibinfo {author} {\bibfnamefont {P.}~\bibnamefont {Torrey}}, \bibinfo {author} {\bibfnamefont {R.}~\bibnamefont {Weinberger}}, \bibinfo {author} {\bibfnamefont {S.}~\bibnamefont {Genel}}, \bibinfo {author} {\bibfnamefont {J.~P.}\ \bibnamefont {Naiman}}, \bibinfo {author} {\bibfnamefont {F.}~\bibnamefont {Marinacci}},\ and\ \bibinfo {author} {\bibfnamefont {M.}~\bibnamefont {Vogelsberger}},\ }\href {https://doi.org/10.1093/mnras/stx3112} {\bibfield  {journal} {\bibinfo  {journal} {Monthly Notices of the Royal Astronomical Society}\ }\textbf {\bibinfo {volume} {475}},\ \bibinfo {pages} {648} (\bibinfo {year} {2017})}\BibitemShut {NoStop}%
\bibitem [{\citenamefont {Naiman}\ \emph {et~al.}(2018)\citenamefont {Naiman}, \citenamefont {Pillepich}, \citenamefont {Springel}, \citenamefont {Ramirez-Ruiz}, \citenamefont {Torrey}, \citenamefont {Vogelsberger}, \citenamefont {Pakmor}, \citenamefont {Nelson}, \citenamefont {Marinacci}, \citenamefont {Hernquist}, \citenamefont {Weinberger},\ and\ \citenamefont {Genel}}]{Naiman_2018}%
  \BibitemOpen
  \bibfield  {author} {\bibinfo {author} {\bibfnamefont {J.~P.}\ \bibnamefont {Naiman}}, \bibinfo {author} {\bibfnamefont {A.}~\bibnamefont {Pillepich}}, \bibinfo {author} {\bibfnamefont {V.}~\bibnamefont {Springel}}, \bibinfo {author} {\bibfnamefont {E.}~\bibnamefont {Ramirez-Ruiz}}, \bibinfo {author} {\bibfnamefont {P.}~\bibnamefont {Torrey}}, \bibinfo {author} {\bibfnamefont {M.}~\bibnamefont {Vogelsberger}}, \bibinfo {author} {\bibfnamefont {R.}~\bibnamefont {Pakmor}}, \bibinfo {author} {\bibfnamefont {D.}~\bibnamefont {Nelson}}, \bibinfo {author} {\bibfnamefont {F.}~\bibnamefont {Marinacci}}, \bibinfo {author} {\bibfnamefont {L.}~\bibnamefont {Hernquist}}, \bibinfo {author} {\bibfnamefont {R.}~\bibnamefont {Weinberger}},\ and\ \bibinfo {author} {\bibfnamefont {S.}~\bibnamefont {Genel}},\ }\href {https://doi.org/10.1093/mnras/sty618} {\bibfield  {journal} {\bibinfo  {journal} {Monthly Notices of the Royal Astronomical Society}\ }\textbf {\bibinfo {volume} {477}},\ \bibinfo {pages} {1206} (\bibinfo {year} {2018})}\BibitemShut {NoStop}%
\bibitem [{\citenamefont {Marinacci}\ \emph {et~al.}(2018)\citenamefont {Marinacci}, \citenamefont {Vogelsberger}, \citenamefont {Pakmor}, \citenamefont {Torrey}, \citenamefont {Springel}, \citenamefont {Hernquist}, \citenamefont {Nelson}, \citenamefont {Weinberger}, \citenamefont {Pillepich}, \citenamefont {Naiman},\ and\ \citenamefont {Genel}}]{Marinacci_2018}%
  \BibitemOpen
  \bibfield  {author} {\bibinfo {author} {\bibfnamefont {F.}~\bibnamefont {Marinacci}}, \bibinfo {author} {\bibfnamefont {M.}~\bibnamefont {Vogelsberger}}, \bibinfo {author} {\bibfnamefont {R.}~\bibnamefont {Pakmor}}, \bibinfo {author} {\bibfnamefont {P.}~\bibnamefont {Torrey}}, \bibinfo {author} {\bibfnamefont {V.}~\bibnamefont {Springel}}, \bibinfo {author} {\bibfnamefont {L.}~\bibnamefont {Hernquist}}, \bibinfo {author} {\bibfnamefont {D.}~\bibnamefont {Nelson}}, \bibinfo {author} {\bibfnamefont {R.}~\bibnamefont {Weinberger}}, \bibinfo {author} {\bibfnamefont {A.}~\bibnamefont {Pillepich}}, \bibinfo {author} {\bibfnamefont {J.}~\bibnamefont {Naiman}},\ and\ \bibinfo {author} {\bibfnamefont {S.}~\bibnamefont {Genel}},\ }\bibfield  {journal} {\bibinfo  {journal} {Monthly Notices of the Royal Astronomical Society}\ }\href {https://doi.org/10.1093/mnras/sty2206} {10.1093/mnras/sty2206} (\bibinfo {year} {2018})\BibitemShut {NoStop}%
\bibitem [{\citenamefont {Vogelsberger}\ \emph {et~al.}(2014{\natexlab{a}})\citenamefont {Vogelsberger}, \citenamefont {Genel}, \citenamefont {Springel}, \citenamefont {Torrey}, \citenamefont {Sijacki}, \citenamefont {Xu}, \citenamefont {Snyder}, \citenamefont {Bird}, \citenamefont {Nelson},\ and\ \citenamefont {Hernquist}}]{vogelsberger2014properties}%
  \BibitemOpen
  \bibfield  {author} {\bibinfo {author} {\bibfnamefont {M.}~\bibnamefont {Vogelsberger}}, \bibinfo {author} {\bibfnamefont {S.}~\bibnamefont {Genel}}, \bibinfo {author} {\bibfnamefont {V.}~\bibnamefont {Springel}}, \bibinfo {author} {\bibfnamefont {P.}~\bibnamefont {Torrey}}, \bibinfo {author} {\bibfnamefont {D.}~\bibnamefont {Sijacki}}, \bibinfo {author} {\bibfnamefont {D.}~\bibnamefont {Xu}}, \bibinfo {author} {\bibfnamefont {G.}~\bibnamefont {Snyder}}, \bibinfo {author} {\bibfnamefont {S.}~\bibnamefont {Bird}}, \bibinfo {author} {\bibfnamefont {D.}~\bibnamefont {Nelson}},\ and\ \bibinfo {author} {\bibfnamefont {L.}~\bibnamefont {Hernquist}},\ }\href@noop {} {\bibfield  {journal} {\bibinfo  {journal} {Nature}\ }\textbf {\bibinfo {volume} {509}},\ \bibinfo {pages} {177} (\bibinfo {year} {2014}{\natexlab{a}})}\BibitemShut {NoStop}%
\bibitem [{\citenamefont {Vogelsberger}\ \emph {et~al.}(2014{\natexlab{b}})\citenamefont {Vogelsberger}, \citenamefont {Genel}, \citenamefont {Springel}, \citenamefont {Torrey}, \citenamefont {Sijacki}, \citenamefont {Xu}, \citenamefont {Snyder}, \citenamefont {Nelson},\ and\ \citenamefont {Hernquist}}]{vogelsberger2014introducing}%
  \BibitemOpen
  \bibfield  {author} {\bibinfo {author} {\bibfnamefont {M.}~\bibnamefont {Vogelsberger}}, \bibinfo {author} {\bibfnamefont {S.}~\bibnamefont {Genel}}, \bibinfo {author} {\bibfnamefont {V.}~\bibnamefont {Springel}}, \bibinfo {author} {\bibfnamefont {P.}~\bibnamefont {Torrey}}, \bibinfo {author} {\bibfnamefont {D.}~\bibnamefont {Sijacki}}, \bibinfo {author} {\bibfnamefont {D.}~\bibnamefont {Xu}}, \bibinfo {author} {\bibfnamefont {G.}~\bibnamefont {Snyder}}, \bibinfo {author} {\bibfnamefont {D.}~\bibnamefont {Nelson}},\ and\ \bibinfo {author} {\bibfnamefont {L.}~\bibnamefont {Hernquist}},\ }\href@noop {} {\bibfield  {journal} {\bibinfo  {journal} {Monthly Notices of the Royal Astronomical Society}\ }\textbf {\bibinfo {volume} {444}},\ \bibinfo {pages} {1518} (\bibinfo {year} {2014}{\natexlab{b}})}\BibitemShut {NoStop}%
\bibitem [{\citenamefont {Genel}\ \emph {et~al.}(2014)\citenamefont {Genel}, \citenamefont {Vogelsberger}, \citenamefont {Springel}, \citenamefont {Sijacki}, \citenamefont {Nelson}, \citenamefont {Snyder}, \citenamefont {Rodriguez-Gomez}, \citenamefont {Torrey},\ and\ \citenamefont {Hernquist}}]{genel2014introducing}%
  \BibitemOpen
  \bibfield  {author} {\bibinfo {author} {\bibfnamefont {S.}~\bibnamefont {Genel}}, \bibinfo {author} {\bibfnamefont {M.}~\bibnamefont {Vogelsberger}}, \bibinfo {author} {\bibfnamefont {V.}~\bibnamefont {Springel}}, \bibinfo {author} {\bibfnamefont {D.}~\bibnamefont {Sijacki}}, \bibinfo {author} {\bibfnamefont {D.}~\bibnamefont {Nelson}}, \bibinfo {author} {\bibfnamefont {G.}~\bibnamefont {Snyder}}, \bibinfo {author} {\bibfnamefont {V.}~\bibnamefont {Rodriguez-Gomez}}, \bibinfo {author} {\bibfnamefont {P.}~\bibnamefont {Torrey}},\ and\ \bibinfo {author} {\bibfnamefont {L.}~\bibnamefont {Hernquist}},\ }\href@noop {} {\bibfield  {journal} {\bibinfo  {journal} {Monthly Notices of the Royal Astronomical Society}\ }\textbf {\bibinfo {volume} {445}},\ \bibinfo {pages} {175} (\bibinfo {year} {2014})}\BibitemShut {NoStop}%
\bibitem [{\citenamefont {Sijacki}\ \emph {et~al.}(2015)\citenamefont {Sijacki}, \citenamefont {Vogelsberger}, \citenamefont {Genel}, \citenamefont {Springel}, \citenamefont {Torrey}, \citenamefont {Snyder}, \citenamefont {Nelson},\ and\ \citenamefont {Hernquist}}]{sijacki2015illustris}%
  \BibitemOpen
  \bibfield  {author} {\bibinfo {author} {\bibfnamefont {D.}~\bibnamefont {Sijacki}}, \bibinfo {author} {\bibfnamefont {M.}~\bibnamefont {Vogelsberger}}, \bibinfo {author} {\bibfnamefont {S.}~\bibnamefont {Genel}}, \bibinfo {author} {\bibfnamefont {V.}~\bibnamefont {Springel}}, \bibinfo {author} {\bibfnamefont {P.}~\bibnamefont {Torrey}}, \bibinfo {author} {\bibfnamefont {G.~F.}\ \bibnamefont {Snyder}}, \bibinfo {author} {\bibfnamefont {D.}~\bibnamefont {Nelson}},\ and\ \bibinfo {author} {\bibfnamefont {L.}~\bibnamefont {Hernquist}},\ }\href@noop {} {\bibfield  {journal} {\bibinfo  {journal} {Monthly Notices of the Royal Astronomical Society}\ }\textbf {\bibinfo {volume} {452}},\ \bibinfo {pages} {575} (\bibinfo {year} {2015})}\BibitemShut {NoStop}%
\bibitem [{\citenamefont {Chen}\ \emph {et~al.}(2023)\citenamefont {Chen}, \citenamefont {Zhang},\ and\ \citenamefont {Yang}}]{chen2023thermal}%
  \BibitemOpen
  \bibfield  {author} {\bibinfo {author} {\bibfnamefont {Z.}~\bibnamefont {Chen}}, \bibinfo {author} {\bibfnamefont {P.}~\bibnamefont {Zhang}},\ and\ \bibinfo {author} {\bibfnamefont {X.}~\bibnamefont {Yang}},\ }\href@noop {} {\bibfield  {journal} {\bibinfo  {journal} {The Astrophysical Journal}\ }\textbf {\bibinfo {volume} {953}},\ \bibinfo {pages} {188} (\bibinfo {year} {2023})}\BibitemShut {NoStop}%
\bibitem [{\citenamefont {Hadzhiyska}\ \emph {et~al.}(2024)\citenamefont {Hadzhiyska}, \citenamefont {Ferraro}, \citenamefont {Guachalla}, \citenamefont {Schaan}, \citenamefont {Aguilar}, \citenamefont {Battaglia}, \citenamefont {Bond}, \citenamefont {Brooks}, \citenamefont {Calabrese}, \citenamefont {Choi} \emph {et~al.}}]{hadzhiyska2024evidence}%
  \BibitemOpen
  \bibfield  {author} {\bibinfo {author} {\bibfnamefont {B.}~\bibnamefont {Hadzhiyska}}, \bibinfo {author} {\bibfnamefont {S.}~\bibnamefont {Ferraro}}, \bibinfo {author} {\bibfnamefont {B.~R.}\ \bibnamefont {Guachalla}}, \bibinfo {author} {\bibfnamefont {E.}~\bibnamefont {Schaan}}, \bibinfo {author} {\bibfnamefont {J.}~\bibnamefont {Aguilar}}, \bibinfo {author} {\bibfnamefont {N.}~\bibnamefont {Battaglia}}, \bibinfo {author} {\bibfnamefont {J.}~\bibnamefont {Bond}}, \bibinfo {author} {\bibfnamefont {D.}~\bibnamefont {Brooks}}, \bibinfo {author} {\bibfnamefont {E.}~\bibnamefont {Calabrese}}, \bibinfo {author} {\bibfnamefont {S.}~\bibnamefont {Choi}}, \emph {et~al.},\ }\href@noop {} {\bibfield  {journal} {\bibinfo  {journal} {arXiv preprint arXiv:2407.07152}\ } (\bibinfo {year} {2024})}\BibitemShut {NoStop}%
\bibitem [{\citenamefont {Guachalla}\ \emph {et~al.}(2025)\citenamefont {Guachalla}, \citenamefont {Schaan}, \citenamefont {Hadzhiyska}, \citenamefont {Ferraro}, \citenamefont {Aguilar}, \citenamefont {Ahlen}, \citenamefont {Battaglia}, \citenamefont {Bianchi}, \citenamefont {Bond}, \citenamefont {Brooks} \emph {et~al.}}]{guachalla2025backlighting}%
  \BibitemOpen
  \bibfield  {author} {\bibinfo {author} {\bibfnamefont {B.~R.}\ \bibnamefont {Guachalla}}, \bibinfo {author} {\bibfnamefont {E.}~\bibnamefont {Schaan}}, \bibinfo {author} {\bibfnamefont {B.}~\bibnamefont {Hadzhiyska}}, \bibinfo {author} {\bibfnamefont {S.}~\bibnamefont {Ferraro}}, \bibinfo {author} {\bibfnamefont {J.~N.}\ \bibnamefont {Aguilar}}, \bibinfo {author} {\bibfnamefont {S.}~\bibnamefont {Ahlen}}, \bibinfo {author} {\bibfnamefont {N.}~\bibnamefont {Battaglia}}, \bibinfo {author} {\bibfnamefont {D.}~\bibnamefont {Bianchi}}, \bibinfo {author} {\bibfnamefont {R.}~\bibnamefont {Bond}}, \bibinfo {author} {\bibfnamefont {D.}~\bibnamefont {Brooks}}, \emph {et~al.},\ }\href@noop {} {\bibfield  {journal} {\bibinfo  {journal} {arXiv preprint arXiv:2503.19870}\ } (\bibinfo {year} {2025})}\BibitemShut {NoStop}%
\bibitem [{\citenamefont {Albuquerque}\ \emph {et~al.}(2025)\citenamefont {Albuquerque}, \citenamefont {Frusciante}, \citenamefont {Sakr}, \citenamefont {Srinivasan}, \citenamefont {Atayde}, \citenamefont {Bose}, \citenamefont {Cardone}, \citenamefont {Casas}, \citenamefont {Martinelli}, \citenamefont {Noller} \emph {et~al.}}]{albuquerque2025euclid}%
  \BibitemOpen
  \bibfield  {author} {\bibinfo {author} {\bibfnamefont {I.}~\bibnamefont {Albuquerque}}, \bibinfo {author} {\bibfnamefont {N.}~\bibnamefont {Frusciante}}, \bibinfo {author} {\bibfnamefont {Z.}~\bibnamefont {Sakr}}, \bibinfo {author} {\bibfnamefont {S.}~\bibnamefont {Srinivasan}}, \bibinfo {author} {\bibfnamefont {L.}~\bibnamefont {Atayde}}, \bibinfo {author} {\bibfnamefont {B.}~\bibnamefont {Bose}}, \bibinfo {author} {\bibfnamefont {V.}~\bibnamefont {Cardone}}, \bibinfo {author} {\bibfnamefont {S.}~\bibnamefont {Casas}}, \bibinfo {author} {\bibfnamefont {M.}~\bibnamefont {Martinelli}}, \bibinfo {author} {\bibfnamefont {J.}~\bibnamefont {Noller}}, \emph {et~al.},\ }\href@noop {} {\bibfield  {journal} {\bibinfo  {journal} {arXiv preprint arXiv:2506.03008}\ } (\bibinfo {year} {2025})}\BibitemShut {NoStop}%
\bibitem [{\citenamefont {Kunz}\ and\ \citenamefont {Sapone}(2007)}]{kunz2007dark}%
  \BibitemOpen
  \bibfield  {author} {\bibinfo {author} {\bibfnamefont {M.}~\bibnamefont {Kunz}}\ and\ \bibinfo {author} {\bibfnamefont {D.}~\bibnamefont {Sapone}},\ }\href@noop {} {\bibfield  {journal} {\bibinfo  {journal} {Physical review letters}\ }\textbf {\bibinfo {volume} {98}},\ \bibinfo {pages} {121301} (\bibinfo {year} {2007})}\BibitemShut {NoStop}%
\bibitem [{\citenamefont {Shi}\ \emph {et~al.}(2024)\citenamefont {Shi}, \citenamefont {Zhang}, \citenamefont {Mao},\ and\ \citenamefont {Gu}}]{shi2024momentum}%
  \BibitemOpen
  \bibfield  {author} {\bibinfo {author} {\bibfnamefont {Y.}~\bibnamefont {Shi}}, \bibinfo {author} {\bibfnamefont {P.}~\bibnamefont {Zhang}}, \bibinfo {author} {\bibfnamefont {S.}~\bibnamefont {Mao}},\ and\ \bibinfo {author} {\bibfnamefont {Q.}~\bibnamefont {Gu}},\ }\href@noop {} {\bibfield  {journal} {\bibinfo  {journal} {Monthly Notices of the Royal Astronomical Society}\ }\textbf {\bibinfo {volume} {528}},\ \bibinfo {pages} {4922} (\bibinfo {year} {2024})}\BibitemShut {NoStop}%
\bibitem [{\citenamefont {Rosselli}\ \emph {et~al.}(2025)\citenamefont {Rosselli}, \citenamefont {Carreres}, \citenamefont {Ravoux}, \citenamefont {Bautista}, \citenamefont {Fouchez}, \citenamefont {Kim}, \citenamefont {Racine}, \citenamefont {Feinstein}, \citenamefont {S{\'a}nchez}, \citenamefont {Valade} \emph {et~al.}}]{rosselli2025forecast}%
  \BibitemOpen
  \bibfield  {author} {\bibinfo {author} {\bibfnamefont {D.}~\bibnamefont {Rosselli}}, \bibinfo {author} {\bibfnamefont {B.}~\bibnamefont {Carreres}}, \bibinfo {author} {\bibfnamefont {C.}~\bibnamefont {Ravoux}}, \bibinfo {author} {\bibfnamefont {J.~E.}\ \bibnamefont {Bautista}}, \bibinfo {author} {\bibfnamefont {D.}~\bibnamefont {Fouchez}}, \bibinfo {author} {\bibfnamefont {A.~G.}\ \bibnamefont {Kim}}, \bibinfo {author} {\bibfnamefont {B.}~\bibnamefont {Racine}}, \bibinfo {author} {\bibfnamefont {F.}~\bibnamefont {Feinstein}}, \bibinfo {author} {\bibfnamefont {B.}~\bibnamefont {S{\'a}nchez}}, \bibinfo {author} {\bibfnamefont {A.}~\bibnamefont {Valade}}, \emph {et~al.},\ }\href@noop {} {\bibfield  {journal} {\bibinfo  {journal} {arXiv preprint arXiv:2507.00157}\ } (\bibinfo {year} {2025})}\BibitemShut {NoStop}%
\bibitem [{\citenamefont {Hui}\ and\ \citenamefont {Greene}(2006)}]{hui2006correlated}%
  \BibitemOpen
  \bibfield  {author} {\bibinfo {author} {\bibfnamefont {L.}~\bibnamefont {Hui}}\ and\ \bibinfo {author} {\bibfnamefont {P.~B.}\ \bibnamefont {Greene}},\ }\href@noop {} {\bibfield  {journal} {\bibinfo  {journal} {Physical Review D—Particles, Fields, Gravitation, and Cosmology}\ }\textbf {\bibinfo {volume} {73}},\ \bibinfo {pages} {123526} (\bibinfo {year} {2006})}\BibitemShut {NoStop}%
\bibitem [{\citenamefont {Pullen}\ \emph {et~al.}(2015)\citenamefont {Pullen}, \citenamefont {Alam},\ and\ \citenamefont {Ho}}]{pullen2015probing}%
  \BibitemOpen
  \bibfield  {author} {\bibinfo {author} {\bibfnamefont {A.~R.}\ \bibnamefont {Pullen}}, \bibinfo {author} {\bibfnamefont {S.}~\bibnamefont {Alam}},\ and\ \bibinfo {author} {\bibfnamefont {S.}~\bibnamefont {Ho}},\ }\href@noop {} {\bibfield  {journal} {\bibinfo  {journal} {Monthly Notices of the Royal Astronomical Society}\ }\textbf {\bibinfo {volume} {449}},\ \bibinfo {pages} {4326} (\bibinfo {year} {2015})}\BibitemShut {NoStop}%
\bibitem [{\citenamefont {Shirasaki}\ \emph {et~al.}(2022)\citenamefont {Shirasaki}, \citenamefont {Takahashi}, \citenamefont {Osato},\ and\ \citenamefont {Ioka}}]{shirasaki2022probing}%
  \BibitemOpen
  \bibfield  {author} {\bibinfo {author} {\bibfnamefont {M.}~\bibnamefont {Shirasaki}}, \bibinfo {author} {\bibfnamefont {R.}~\bibnamefont {Takahashi}}, \bibinfo {author} {\bibfnamefont {K.}~\bibnamefont {Osato}},\ and\ \bibinfo {author} {\bibfnamefont {K.}~\bibnamefont {Ioka}},\ }\href@noop {} {\bibfield  {journal} {\bibinfo  {journal} {Monthly Notices of the Royal Astronomical Society}\ }\textbf {\bibinfo {volume} {512}},\ \bibinfo {pages} {1730} (\bibinfo {year} {2022})}\BibitemShut {NoStop}%
\bibitem [{\citenamefont {Madhavacheril}\ \emph {et~al.}(2019)\citenamefont {Madhavacheril}, \citenamefont {Battaglia}, \citenamefont {Smith},\ and\ \citenamefont {Sievers}}]{madhavacheril2019cosmology}%
  \BibitemOpen
  \bibfield  {author} {\bibinfo {author} {\bibfnamefont {M.~S.}\ \bibnamefont {Madhavacheril}}, \bibinfo {author} {\bibfnamefont {N.}~\bibnamefont {Battaglia}}, \bibinfo {author} {\bibfnamefont {K.~M.}\ \bibnamefont {Smith}},\ and\ \bibinfo {author} {\bibfnamefont {J.~L.}\ \bibnamefont {Sievers}},\ }\href@noop {} {\bibfield  {journal} {\bibinfo  {journal} {Physical Review D}\ }\textbf {\bibinfo {volume} {100}},\ \bibinfo {pages} {103532} (\bibinfo {year} {2019})}\BibitemShut {NoStop}%
\bibitem [{\citenamefont {Sharma}\ \emph {et~al.}(2025)\citenamefont {Sharma}, \citenamefont {Krause}, \citenamefont {Ravi}, \citenamefont {Reischke}, \citenamefont {Connor}, \citenamefont {Anbajagane} \emph {et~al.}}]{sharma2025probing}%
  \BibitemOpen
  \bibfield  {author} {\bibinfo {author} {\bibfnamefont {K.}~\bibnamefont {Sharma}}, \bibinfo {author} {\bibfnamefont {E.}~\bibnamefont {Krause}}, \bibinfo {author} {\bibfnamefont {V.}~\bibnamefont {Ravi}}, \bibinfo {author} {\bibfnamefont {R.}~\bibnamefont {Reischke}}, \bibinfo {author} {\bibfnamefont {L.}~\bibnamefont {Connor}}, \bibinfo {author} {\bibfnamefont {D.}~\bibnamefont {Anbajagane}}, \emph {et~al.},\ }\href@noop {} {\bibfield  {journal} {\bibinfo  {journal} {arXiv preprint arXiv:2509.05866}\ } (\bibinfo {year} {2025})}\BibitemShut {NoStop}%
\bibitem [{\citenamefont {Smith}\ and\ \citenamefont {Ferraro}(2017)}]{smith2017detecting}%
  \BibitemOpen
  \bibfield  {author} {\bibinfo {author} {\bibfnamefont {K.~M.}\ \bibnamefont {Smith}}\ and\ \bibinfo {author} {\bibfnamefont {S.}~\bibnamefont {Ferraro}},\ }\href@noop {} {\bibfield  {journal} {\bibinfo  {journal} {Physical Review Letters}\ }\textbf {\bibinfo {volume} {119}},\ \bibinfo {pages} {021301} (\bibinfo {year} {2017})}\BibitemShut {NoStop}%
\bibitem [{\citenamefont {Kumar}\ \emph {et~al.}(2025)\citenamefont {Kumar}, \citenamefont {{\c{C}}al{\i}{\c{s}}kan}, \citenamefont {Hotinli}, \citenamefont {Smith},\ and\ \citenamefont {Kamionkowski}}]{kumar2025electrons}%
  \BibitemOpen
  \bibfield  {author} {\bibinfo {author} {\bibfnamefont {N.~A.}\ \bibnamefont {Kumar}}, \bibinfo {author} {\bibfnamefont {M.}~\bibnamefont {{\c{C}}al{\i}{\c{s}}kan}}, \bibinfo {author} {\bibfnamefont {S.~C.}\ \bibnamefont {Hotinli}}, \bibinfo {author} {\bibfnamefont {K.}~\bibnamefont {Smith}},\ and\ \bibinfo {author} {\bibfnamefont {M.}~\bibnamefont {Kamionkowski}},\ }\href@noop {} {\bibfield  {journal} {\bibinfo  {journal} {arXiv preprint arXiv:2509.18249}\ } (\bibinfo {year} {2025})}\BibitemShut {NoStop}%
\bibitem [{\citenamefont {Driver}(2021)}]{driver2021challenge}%
  \BibitemOpen
  \bibfield  {author} {\bibinfo {author} {\bibfnamefont {S.}~\bibnamefont {Driver}},\ }\href@noop {} {\bibfield  {journal} {\bibinfo  {journal} {Nature Astronomy}\ }\textbf {\bibinfo {volume} {5}},\ \bibinfo {pages} {852} (\bibinfo {year} {2021})}\BibitemShut {NoStop}%
\bibitem [{\citenamefont {Fukugita}\ and\ \citenamefont {Peebles}(2004)}]{fukugita2004cosmic}%
  \BibitemOpen
  \bibfield  {author} {\bibinfo {author} {\bibfnamefont {M.}~\bibnamefont {Fukugita}}\ and\ \bibinfo {author} {\bibfnamefont {P.~J.~E.}\ \bibnamefont {Peebles}},\ }\href@noop {} {\bibfield  {journal} {\bibinfo  {journal} {The Astrophysical Journal}\ }\textbf {\bibinfo {volume} {616}},\ \bibinfo {pages} {643} (\bibinfo {year} {2004})}\BibitemShut {NoStop}%
\bibitem [{\citenamefont {Blas}\ \emph {et~al.}(2011)\citenamefont {Blas}, \citenamefont {Lesgourgues},\ and\ \citenamefont {Tram}}]{blas2011cosmic}%
  \BibitemOpen
  \bibfield  {author} {\bibinfo {author} {\bibfnamefont {D.}~\bibnamefont {Blas}}, \bibinfo {author} {\bibfnamefont {J.}~\bibnamefont {Lesgourgues}},\ and\ \bibinfo {author} {\bibfnamefont {T.}~\bibnamefont {Tram}},\ }\href@noop {} {\bibfield  {journal} {\bibinfo  {journal} {Journal of Cosmology and Astroparticle Physics}\ }\textbf {\bibinfo {volume} {2011}}\bibinfo  {number} { (07)},\ \bibinfo {pages} {034}}\BibitemShut {NoStop}%
\bibitem [{\citenamefont {Angulo}\ \emph {et~al.}(2013)\citenamefont {Angulo}, \citenamefont {Hahn},\ and\ \citenamefont {Abel}}]{angulo2013closely}%
  \BibitemOpen
\bibfield  {number} {  }\bibfield  {author} {\bibinfo {author} {\bibfnamefont {R.~E.}\ \bibnamefont {Angulo}}, \bibinfo {author} {\bibfnamefont {O.}~\bibnamefont {Hahn}},\ and\ \bibinfo {author} {\bibfnamefont {T.}~\bibnamefont {Abel}},\ }\href@noop {} {\bibfield  {journal} {\bibinfo  {journal} {Monthly Notices of the Royal Astronomical Society}\ }\textbf {\bibinfo {volume} {434}},\ \bibinfo {pages} {1756} (\bibinfo {year} {2013})}\BibitemShut {NoStop}%
\bibitem [{\citenamefont {Khoraminezhad}\ \emph {et~al.}(2021)\citenamefont {Khoraminezhad}, \citenamefont {Lazeyras}, \citenamefont {Angulo}, \citenamefont {Hahn},\ and\ \citenamefont {Viel}}]{khoraminezhad2021quantifying}%
  \BibitemOpen
  \bibfield  {author} {\bibinfo {author} {\bibfnamefont {H.}~\bibnamefont {Khoraminezhad}}, \bibinfo {author} {\bibfnamefont {T.}~\bibnamefont {Lazeyras}}, \bibinfo {author} {\bibfnamefont {R.~E.}\ \bibnamefont {Angulo}}, \bibinfo {author} {\bibfnamefont {O.}~\bibnamefont {Hahn}},\ and\ \bibinfo {author} {\bibfnamefont {M.}~\bibnamefont {Viel}},\ }\href@noop {} {\bibfield  {journal} {\bibinfo  {journal} {Journal of Cosmology and Astroparticle Physics}\ }\textbf {\bibinfo {volume} {2021}}\bibinfo  {number} { (03)},\ \bibinfo {pages} {023}}\BibitemShut {NoStop}%
\bibitem [{\citenamefont {Chen}\ \emph {et~al.}(2025)\citenamefont {Chen}, \citenamefont {Yu}, \citenamefont {Han},\ and\ \citenamefont {Jing}}]{chen2025csst}%
  \BibitemOpen
\bibfield  {number} {  }\bibfield  {author} {\bibinfo {author} {\bibfnamefont {Z.}~\bibnamefont {Chen}}, \bibinfo {author} {\bibfnamefont {Y.}~\bibnamefont {Yu}}, \bibinfo {author} {\bibfnamefont {J.}~\bibnamefont {Han}},\ and\ \bibinfo {author} {\bibfnamefont {Y.}~\bibnamefont {Jing}},\ }\href@noop {} {\bibfield  {journal} {\bibinfo  {journal} {Science China Physics, Mechanics \& Astronomy}\ }\textbf {\bibinfo {volume} {68}},\ \bibinfo {pages} {289512} (\bibinfo {year} {2025})}\BibitemShut {NoStop}%
\end{thebibliography}%

\newpage
\onecolumngrid

\appendix

\section{Cosmic baryon budget and clustering bias}
\label{sec:appendix}

The baryonic density contrast can be decomposed as 
\begin{equation}
\delta_b = f_{\rm HII} \delta_{\rm HII} + f_{\rm HeIII} \delta_{\rm HeIII}
+ f_{*} \delta_{*} 
+ f_{\rm HI} \delta_{\rm HI} + f_{\rm HeI} \delta_{\rm HeI} + 
\sum_i f_{Z_i} \delta_{Z_i} 
\end{equation}
where $\delta_i \equiv \rho_i/\bar{\rho}_i -1$ is the mass density contrast, and $f_i\equiv\Omega_i/\Omega_b$ denotes the mass fraction of $i$-component relative to the total baryon. 
In the low redshift $z\lesssim 2$, the cosmic baryon is fully ionized, and the dominant baryon budgets are fully ionized hydrogen and helium in the warm and hot plasma, accounting for a fraction of $f_{\rm HII} +f_{\rm HeIII} \sim 0.9$. The subdominant components are stars and stellar remnants with $f_{*}\sim 0.05$, and the neutral gas with $f_{\rm HI} + f_{\rm HeI}\sim 0.05$ \cite{driver2021challenge, connor2025gas}. 
The other components $Z_i$, such as metals, may be important tracers in the detection of baryon distribution, but they are rare in the baryon budget \cite{fukugita2004cosmic}. Thus, we consider the total cosmic baryons consisting of three components: HII+HeIII, stellar contents, and neutral gas traced by HI. Moreover, the electrical neutrality leads to $ \delta_{\rm HII} \simeq \delta_{\rm HeIII} \simeq \delta_e $. We therefore obtain 
\begin{equation}\label{equ:delta_b_into_others}
\delta_b  \simeq  
\left( f_{\rm HII} + f_{\rm HeIII} \right) \delta_e
+ f_{*} \delta_{*}
+ {X_{\rm H}+X_{\rm He}\over X_{\rm H} } f_{\rm HI} \delta_{\rm HI}  
\end{equation}
The clustering bias $b_i \equiv P_{im}/P_{mm}$ is defined relative to the total matter fluctuation $\delta_m = ( \Omega_c\delta_c + \Omega_b\delta_b) /\Omega_m$ in the linear region, where $\delta_c$ indicates the cold dark matter perturbation. 
On sub-horizon scales prior to the period of recombination, baryons are tightly coupled to photons through Compton scattering, while the non-interacting dark matter only senses the gravitational force. It results in distinct evolutions of baryon and dark matter perturbations. 
After recombination, gravity governs the evolution of matter components on large scales, and the equivalence principle ensures that the total baryon component and dark matter experience the same acceleration when falling into gravitational potential wells. 
The co-motion erases the differences between baryon and dark matter perturbations, and makes $\delta_b$ an unbiased tracer of $\delta_c$ or equivalent $\delta_m$. 
Within the scales $k \lesssim 0.1\,h/{\rm Mpc}$ and redshifts $0\lesssim z\lesssim 2$ of interest, both the linear perturbation by Boltzmann solver like \texttt{CLASS} \cite{blas2011cosmic} and the nonlinear evolution in 2-fluid simulations including dark matter and baryon \cite{angulo2013closely, khoraminezhad2021quantifying} have demonstrated that the relative difference between $\delta_b$ and $\delta_c$ is confined to the sub-percent level on large scales. 
Moreover, this also justifies the common assumption in gravity-only simulations that treat the combination of baryon and dark matter as a single fluid during late-time evolution \cite{chen2025csst}.
Therefore, we assume the baryon bias as $b_b=1$ and obtain an estimation of the electron bias,  
\begin{equation} \label{equ:be_indirect}
b_e  = {1\over f_{\rm HII} + f_{\rm HeIII} }  
\left(  1 - f_{*} b_{*} 
- {X_{\rm H}+X_{\rm He}\over X_{\rm H} } f_{\rm HI} b_{\rm HI}  \right)   \;.     \\
\end{equation}
where $b_e=1$ if there are $b_{*}=1$ and $b_{\rm HI}=1$. 

The spatial inhomogeneity of the free electron distribution is
\begin{equation} \label{equ:ne_express}
n_e - \bar{n}_e = {\bar\rho_c\,\Omega_b\over m_p} \left( f_{\rm HII} + {1\over 2}f_{\rm HeIII} + \sum_{i} N_{i} f_{Z_i^+} \right) \,\delta_e   \quad,
\end{equation}
where $f_{Z_i^+}$ denotes the mass fraction of ionized gas other than HII and HeIII, and $N_i$ represents the number of ionized electrons contributed per proton/neutron, e.g., $N_{\rm HeIII}=1/2$. Because of $f_{Z_i^+} \ll f_{\rm HII} \sim f_{\rm HeIII}$, we obtain 
\begin{align}
n_e - \bar{n}_e  
&=  {\bar\rho_c\,\Omega_b\over m_p} 
\left( f_{\rm HII} + {1\over 2}f_{\rm HeIII}  \right) \,\delta_e   
\label{equ:ne_express_fe}  \\
&\simeq  {\bar\rho_c\,\Omega_b\over m_p} 
{ X_{\rm H} + {1\over 2} X_{\rm He}  \over  X_{\rm H} + X_{\rm He}  }
\left( 1  - f_{*} b_{*} 
- {X_{\rm H}+X_{\rm He}\over X_{\rm H} } f_{\rm HI} b_{\rm HI}  \right) \,\delta_m    
\label{equ:ne_express_be}
\end{align}
where the second equality follows the relation ${f_{\rm HII} / f_{\rm HeIII}} \simeq {X_{\rm H}/ X_{\rm He}}$. $X_{\rm H}\simeq 0.76$ and $X_{\rm He}\simeq 0.24$ are mass abundances of hydrogen and helium elements. 
By definition, we can obtain the expression for $f_eb_e$ of Eq.~(\ref{equ:be_from_bb}) in the main text. 

\begin{figure}
\includegraphics[width=0.90\columnwidth]{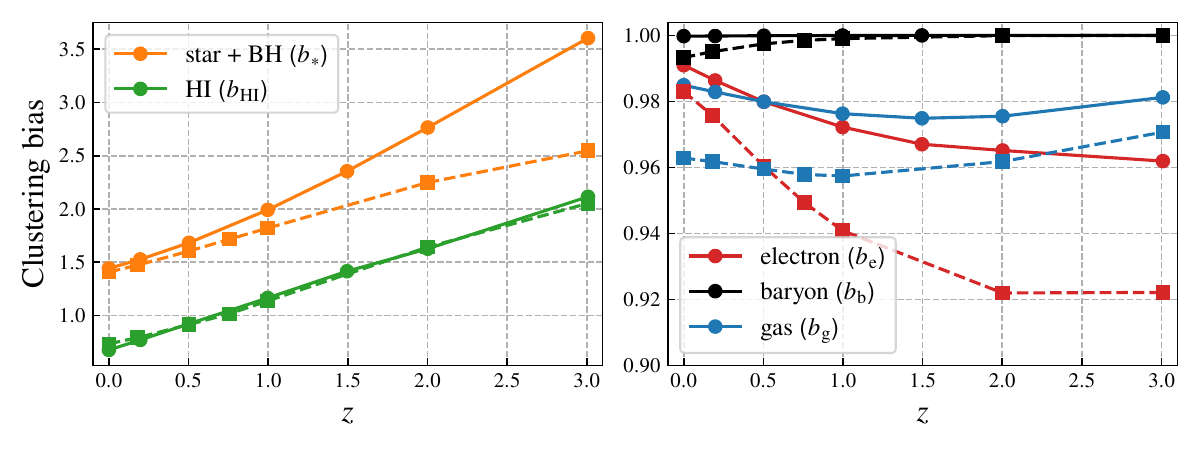}
\caption{ \label{fig:tracer_bias} 
The clustering bias $b_i=P_{im}/P_{mm}$ of $i$-species component measured in TNG300-1 (solid lines) and Illustris-1 (dashed lines) simulations. 
The \textit{left panel} shows the bias measured for stars and black holes (orange lines) and for neutral hydrogen (green lines). 
Since both stars and black holes form in overdense regions of the cosmic web, their bias values are typically greater than unity. A similar trend holds for neutral hydrogen at early times, but astrophysical processes deplete neutral gas in massive halos, leading to a decline in its bias value at later times. 
The \textit{right panel} shows the bias measured for electrons (red lines), total baryons (black lines), and gas components (blue lines). The apparent deviation $b_b\lesssim 1$ at low redshifts arises from the extremely strong AGN feedback implemented in the Illustris-1 simulation, and its potential impact is already considered in the simulation validation presented in the main text. 
These measurements are consistent with the results presented in the IllustrisTNG publication \cite{Springel_2017}. 
}
\end{figure}

\begin{figure}
\includegraphics[width=0.90\columnwidth]{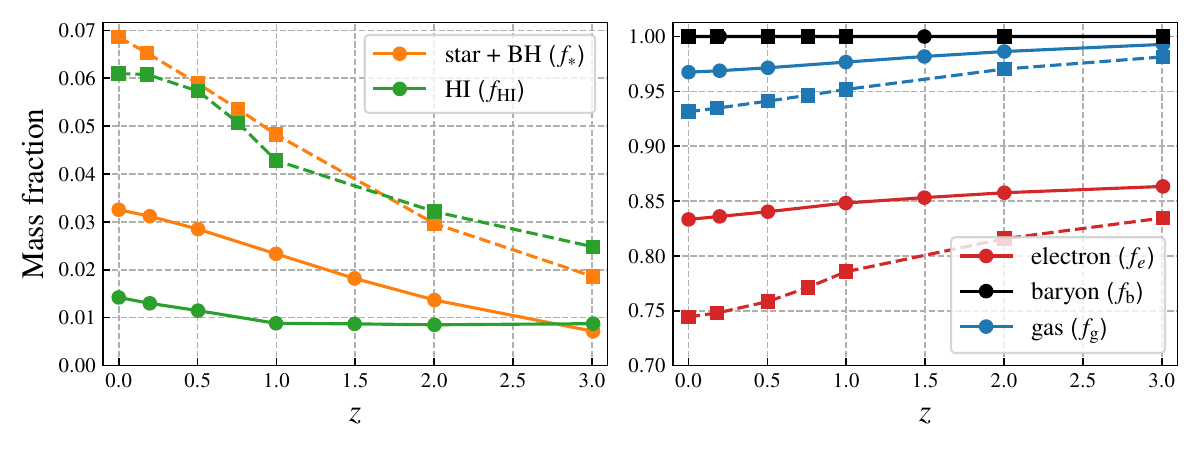}
\caption{ \label{fig:tracer_fraction} 
Baryon mass fraction $f_i = \Omega_i/\Omega_b$ measured in simulations. The labels of components are the same as Fig.~\ref{fig:tracer_bias}. The electron fraction is given by $f_e= f_{\rm HII} + {1\over 2}f_{\rm HeIII} \simeq  (X_{\rm H} + {1\over 2} X_{\rm He})\, f_{\rm HII}/f_{\rm H}$, where $f_{\rm HII}$, $X_{\rm H}$ and $X_{\rm He}$ are directly accessed in simulation products. 
The large differences in the cold gas fractions between TNG300-1 and Illustris-1 indicate that these two simulation suites adopt highly distinct subgrid physics.
}
\end{figure}

In the main text, we argue that the electron bias can be inferred using the relation Eq.~(\ref{equ:be_from_bb}), or equivalently Eq.~(\ref{equ:be_indirect}), hence alleviating the systematic bias. 
This approximation is validated in Fig.~\ref{fig:electron_bias}, using hydrodynamical simulations TNG300-1 \cite{Springel_2017, Nelson_2017, Pillepich_2017, Naiman_2018, Marinacci_2018} and Illustris-1 \cite{vogelsberger2014properties, vogelsberger2014introducing, genel2014introducing, sijacki2015illustris}.
These simulations impose the same initial conditions for dark matter and baryon particles at $z=127$, and the baryon components are further separated into three species: gas, star, and black hole, during the subsequent evolution. 
In Fig.~\ref{fig:tracer_bias} and Fig.~\ref{fig:tracer_fraction}, we present the measurements of clustering bias and mass fraction for different baryonic components. The clustering bias is estimated by $\hat{b}_i = \frac{\sum_{k<k_{\rm max}} \hat{P}_{im}(k)}{\sum_{k<k_{\rm max}} \hat{P}_{mm}(k)}$, with a scale cut $k_{\rm max}=0.12\,h/{\rm Mpc}$ for both simulation results. Instead of using $k_{\rm max}=0.10\,h/{\rm Mpc}$ in the forecast, we use a slightly larger value to mitigate the sample variance in the Illustris-1 simulation due to the limited box size $L=75\,h^{-1}{\rm Mpc}$.

The bias of the total baryon is expected to be unity, i.e., $b_b=1$, as implied by simulation setups and the equivalence principle. However, extremely strong baryonic feedback may affect large-scale clustering of baryons up to Mpc scales, particularly at low redshift. Such non-gravitational forces can redistribute matter and induce deviations from unit baryon bias assumption, thereby degrading the accuracy of the systematic mitigation using Eq.~(\ref{equ:be_from_bb}). 
The validation results from Illustris-1 provide a conservative benchmark for these potential small-scale effects. Here, the baryonic effects in Illustris suite are widely recognized as one of the strongest feedback implementations among the modern hydrodynamical simulations \cite{schaller2025flamingo}. 
As shown in Fig.~\ref{fig:tracer_bias}, the assumption $b_b=1$ is well satisfied in TNG300-1, but not in Illustris-1. In the latter, the extremely strong baryonic feedback suppresses the baryon clustering amplitude significantly and impacts scales up to $k\sim 0.1\,h/{\rm Mpc}$, leading to the deviation $1 - \hat{b}_b \sim 0.5\%$ at $z\lesssim 0.5$. 
This deviation propagates into the validation of Eq.~(\ref{equ:be_from_bb}), where it contributes an amount of $\sim 0.8\%$ to the total deviation $| \widehat{f_eb}_e / (f_eb_e) -1 | \sim 1.3\%$ at $z\lesssim 0.3$, dominating the residual systematic at low redshifts. Consequently, strong baryonic feedback partially accounts for the significantly larger residual systematic in Illustris-1 ($\sim 1.2\%$) compared to TNG300-1 ($\sim 0.2\%$) at $z\lesssim 0.3$. 
Nevertheless, on the one hand, the simulation results indicate that $k_{\max}\simeq 0.1\,h/{\rm Mpc}$ is sufficient to achieve $\sim 1\%$ accuracy in Illustris-like feedback scenarios. On the other hand, these baryonic effects are expected to vanish if a more conservative scale cut is applied, while the statistical significance is not substantially degraded (e.g., comparison between $k_{\rm max}= 0.05\, h/{\rm Mpc}$ and the fiducial $k_{\rm max}= 0.1\, h/{\rm Mpc}$ cut in the bottom panel of Fig.~\ref{fig:FG}). Therefore, for baryonic effects under reasonable expectation, i.e., not significantly stronger than those in Illustris-1, our conclusions remain unchanged.

In Fig.~\ref{fig:tracer_fraction}, we present the measurement of the mass fraction of different components. Though the fraction of $f_{*}$ or $f_{\rm HI}$ is small at high redshift $z\gtrsim 1$, their bias values are significantly larger than unity, also leading to a suppression of the electron clustering according to Eq.~(\ref{equ:be_indirect}).

Because the diffuse gas preferentially resides in the underdense regions of the cosmic web, the clustering of free electrons is expected to be suppressed relative to the total matter field, leading to $b_e < 1$. Thus, the neglect of electron bias would overestimate the $F_G$ value by $F_G\propto b_e^{-1}$. 
Particularly, one can understand the fact of $b_e<1$ by the relation Eq.~(\ref{equ:delta_b_into_others}) or Eq.~(\ref{equ:be_indirect}). In the leading order of $\mathcal{O}(f_i)$, the electron bias is 
\begin{equation}\label{equ:be_linear}
b_e \simeq 
1 - f_{*} \left( b_{*} -1 \right)
- {X_{\rm H}+X_{\rm He}\over X_{\rm H} } f_{\rm HI} \left(b_{\rm HI} -1 \right)   \;,
\end{equation}
from which the electron bias generally deviates from $b_e\simeq 1$ by an amount of order $f_ib_i\sim \mathcal{O}(10^{-2})$. 
The stars and stellar remnants form in the overdense regions of the cosmic web, so their bias values are typically greater than unity, $b_{*}>1$. 
On the other hand, neutral hydrogens are also bound within halos, but depleted in massive halos due to astrophysical processes, leading to $b_{\rm HI}> 1$ at early times but $b_{\rm HI}\lesssim 1$ in late times. 
Typically, there are $f_{*}\sim f_{\rm HI}$ and $b_{*}>b_{\rm HI}\gtrsim 1$, where the latter relation also reflects the fact that star formation preferentially occurs in the dense cold gas. As a consequence, in Eq.~(\ref{equ:be_linear}), the second term dominates over the third term, and the summation of these two terms is therefore expected to be positive. Given that the baryon bias satisfies $b_b=1$ on scales dominated by gravity, the electron bias is suggested to be $b_e<1$. These inferences are also supported by the simulation results shown in Fig.~\ref{fig:tracer_bias} and Fig.~\ref{fig:tracer_fraction}.




\end{document}